\documentclass[showpacs,aps,pre,twocolumn]{revtex4}
\usepackage{graphicx}
\usepackage{amsfonts}

\begin{document}
\title{Localized modes of binary mixtures of Bose-Einstein condensates \\
in nonlinear optical lattices}
\author{F. Kh. Abdullaev$^{1,2}$, A. Gammal$^{3}$,   M. Salerno $^{2,4}$, Lauro Tomio$^{2}$}
\affiliation{$^{1}$ Physical-Technical Institute of the  Academy
of Sciences, Tashkent, Uzbekistan
\\
$^2$Instituto de F\'\i sica Te\'orica, UNESP, Rua Pamplona, 145,
01405-900, S\~ao Paulo, SP, Brasil
\\
$^3$ Instituto de F\'isica, Universidade de S\~ao Paulo,
05315-970, C.P. 66318, S\~ao Paulo, SP, Brasil
\\
$^4$ Dipartimento di Fisica ``E. R. Caianiello", Consorzio
Nazionale Interuniversitario per le Scienze Fisiche della Materia
(CNISM), Universit\'{a} di Salerno, I-84081, Baronissi (SA),
Italy}

\begin{abstract}
The properties of the localized states of a two component
Bose-Einstein condensate confined in a nonlinear periodic
potential [nonlinear optical lattice] are investigated. We reveal
the existence of new types of solitons and study their stability
by means of analytical and numerical approaches. The symmetry
properties of the localized states with respect to the NOL are
also investigated. We show that nonlinear optical lattices allow
the existence of bright soliton modes with equal symmetry in both
components, bright localized modes of mixed symmetry type, as well
as, dark-bright bound states and bright modes on periodic
backgrounds. In spite of the quasi 1D nature of the problem,  the
fundamental symmetric localized modes undergo a delocalizing
transition when the strength of the nonlinear optical lattice is
varied. This transition is associated with the existence of an
unstable solution, which exhibits a shrinking (decaying) behavior
for slightly overcritical (undercritical) variations in the number
of atoms.

\end{abstract}
\date{\today}
\maketitle

\section{Introduction}
Bose-Einstein condensates (BEC)  in optical lattices (OL) have
recently attracted a great deal of attention due to the
possibility of investigating, both at the theoretical and at the
experimental level, interesting physical phenomena
such as Bloch oscillations, Landau Zener tunneling, Mott
transitions, etc. \cite{Morsh,BK}.

The interplay between the nonlinearity (intrinsic in the
interatomic interaction) and the periodic structure (induced by
the OL) leads to the formation of localized states trough the
mechanism of  the modulational instability of the Bloch states at
the edges of the Brillouin zone  of the underlying linear periodic
system \cite{KS02}. These states, also known as gap-solitons, can
exist in presence of both attractive and repulsive interactions
\cite{TS,ABDKS,Carus}, this last fact being possible only due to
the presence of the OL.

The existence of gap solitons in repulsive BEC was experimentally
demonstrated in \cite{Eier}. The phenomena of Bloch oscillations,
generation of coherent atomic pulses (atom laser)\cite{Kas},
superfluid-Mott transition\cite{Gren}, were also experimentally
observed. The OLs considered in these experiments act as external
potentials (and therefore linearly) on the condensate, this
introducing an intrinsic (state independent) periodicity  in the
system. In the following we shall refer to this type of lattices
as {\it linear} OLs  (LOLs). In higher dimensions LOLs were shown
to be very effective in stabilizing localized states against
collapse or decay, leading to the formation of stable
multidimensional solitons \cite{BKS02}.

Besides LOLs, it is also possible to consider {\it nonlinear} OLs
(NOLs) with symmetry properties which depend on the  wavefunction
characterizing  the state of the system. A NOL can be obtained by
inducing a periodic spatial variation of the two body interatomic
interaction strength (atomic scattering length), leading to a
periodic space modulation of the nonlinear coefficient  in the
Gross-Pitaevskii equation (GPE) governing the mean field dynamics
of the ground state. This periodic modulation can be
experimentally achieved either by means of the standard Feshbach
resonance method \cite{Inouye}, taking an external magnetic field
near the resonance which is spatially periodic
\cite{AS03,kevrekidis,AGKT,GA,Niar}, or by the optically induced
Feshbach resonance technique. In the last case the nonlinear
periodic potential can be produced by two counter propagating
laser beams with parameters near the optically induced Feshbach
resonance \cite{SM,AG}. A periodic variation of the laser field
intensity in space and a proper choice of the resonance detuning
lead to a spatial dependence of the scattering length \cite{OFR2}
and hence to a spatial dependent nonlinear coefficient in the GPE.

Different interesting phenomena occurring in BEC in presence of a
NOL have already been studied, such as the transmission of
wave packets through nonlinear barriers, generation of atomic
solitons, and existence of localized states
Refs.~\cite{SM,AG,Konotop06,AAG}. Mathematical properties of the
ground state and the existence of localized states of quasi-1D BEC
in NOL have also been studied in \cite{Fibich,Garcia}. All these
studies \cite{AAG,Bambi,Bludov} concern mainly with scalar (single
component) 1D BEC in NOL . The possibility of stabilizing
multi-dimensional scalar solitons by means of NOLs is presently
under investigation (preliminary studies show that NOLs are unable
to stabilize 2D solitons if the average nonlinearity is negative),
while multi-component BECs in NOL have not been considered yet
neither theoretically nor experimentally. This last problem arises
when two or more BEC atomic species interact in presence of
periodic spatial modulations of the  scattering lengths, which
can occur between the species (inter-species) and/or 
within the species (intra-species). 
The interaction between the two BEC components leads to an inter-species 
NOL which can play a stabilizing role for localized states. 
Spatial modulations of the intra-species scattering length (giving rise 
to intra-species NOLs) can also lead to the existence of new types of 
soliton states.

The aim of the present paper is to study the properties of the
localized states of  two-component BEC mixtures in NOLs. The case
of a sinusoidal variation in space of the intra- and inter-species
scattering lengths will be considered. In particular we show the
existence of new types of solitons and study their stability by
means of  analytical and numerical methods. The symmetry
properties of the localized with respect to the NOL are also
investigated. We show that the  NOL allows the existence of bright
soliton modes with equal symmetry in both components, bright
localized modes of mixed symmetry type, as well as bright-dark
bound states and bright modes on periodic backgrounds. We also
show that, in spite of the quasi 1D nature of the problem, the
fundamental symmetric localized modes undergo a delocalizing
transition when the strength of the non linear optical lattice is
varied. This transition is associated with the existence of an
unstable solution  which exhibit a shrinking (decaying) behavior
for slightly overcritical (undercritical) variations in the number
of atoms.

The phenomenon of the delocalizing transition was also
investigated in \cite{BS04} for the case of  multidimensional
single component BEC solitons in LOL and in \cite{Bludov} for the
case of one-dimensional BEC's with combined linear and nonlinear
OL's. Delocalizing transitions in binary BEC mixtures have not
been previously investigated.

For  the analysis of strongly localized modes (i.e. localized in
one or few cells of the NOL) we will apply the variational
approach which was shown to be effective for such type of
problems, while for delocalizing transitions and broad solitons we
use a vectorial Gross-Pitaevskii equation averaged over rapid
variations in space of the nonlinear potential. Results are then
compared with those obtained by direct numerical simulations of
the coupled GPE system. As numerical tools to investigate the
above problems we use both self-consistent exact diagonalizations
\cite{LP05} and generalized relaxing methods \cite{marijana}.

The paper is organized as follows. In Section II  we describe the
physical model for the two component BEC under  action of a NOL,
based on the optical manipulation of the scattering length by
optically  induced Feshbach resonances. The model equations are
introduced in the mean field approximation in terms of  two
coupled 1D Gross-Pitaevskii equations with intra- and
inter-species interaction terms. The problem of existence of
soliton solutions (when the inter- and intra-species atomic
scattering lengths are periodically modulated in space), the
symmetry properties of localized modes and their stability are 
discussed in Sections III. The delocalizing transitions of
fundamental modes and the existence of unstable solutions
associated with them are studied in section IV. The analytical
predictions are confirmed by direct numerical simulations of the
full GP equation (Sections II-IV). Finally, in Section V, the main
results of the paper are summarized.

\section{The model}
Two component condensate represent the mixture of atoms in the
different hyperfine states\cite{Mies,Simoni,KT,Kevrekidis2}. We
consider here the dynamics of two-component BEC in presence of a
nonlinear optical lattice produced either by spatially varying
magnetic fields near a Feshbach resonance (FR) value or by
optically induced Feshbach resonances \cite{OFR2}. According to
this last approach, the scattering length $a_s$ can be optically
manipulated if the incident light is close to the resonance with
one of the bound $p$ levels of electronically excited molecules.
Virtual radiative transitions of a pair of interacting atoms to
this level, can change value and/or reverse the sign of the
scattering length. The periodic variation of the laser field
intensity in the standing wave, $I(x) = I_{0}\cos^{2}(kx)$, produces
periodic variation of the atomic scattering length,  such that
\begin{equation}\label{ax}
a_{s}(x) = a_{s0}\left[1 + \alpha \frac{I}{\delta + I}\right],
\end{equation}
where $a_{s0}$ is the scattering length in the absence of light,
$\delta$ is the frequency detuning of the light 
from the FR and $\alpha $ is a constant factor.
 For weak intensities, when $I_0 \ll |\delta|$, we have that  
 $a_{s} = a_{s0} + a_{s1}\cos^{2}(kx) $.
Periodic variation of the scattering length by a spatially
varying external magnetic field ${\cal B}(x)$ near a Feshbach
resonance (FR), can be described by
\begin{equation}
a_{s}(x) = a_{s0}\left(1 + \frac{\Delta}{{\cal B}_0 - {\cal B}(x)}\right),
\end{equation}
where 
${\cal B}_0$ is the resonant value and $\Delta$ the corresponding width. 
Examples are:
a multicomponent BEC of $^{23}$Na atoms\cite{Mies} or the mixture
of $^{41}$K - $^{87}$Rb atoms on the surface of a chip. The
periodic variation of ${\cal B}$ can be controlled by the current
in a magnetic wire on the chip surface\cite{Gimp}. For the mixture
$^{41}$K - $^{87}$Rb it was shown recently that the inter-species
scattering length $a_{12}$ can be tuned using the Feshbach
resonances by varying the external magnetic field in the interval
$(50-800)$G \cite{Simoni}.

The mean field equations for the ground state wavefunction of a
quasi-1D two-component BEC under the action of a NOL are given by
the following coupled GP equations \cite{SM,AG}:
  \begin{eqnarray}
{\rm i}\hbar\frac{\partial \psi_i}{\partial \bar{t}} =
-\frac{\hbar^2}{2m}
   \frac{\partial^2 \psi_i}{\partial \bar{x}^2} + B_{i}(\bar{x})|\psi_i|^2
   \psi_i + S_{12}(\bar{x})|\psi_{3-i}|^2\psi_i,
\end{eqnarray}
where $i=1,2$ refer to the component index, and the
full-dimensional space and time variables are given by $\bar x$
and $\bar t$. In the above, $S_{12}$ is the parameter giving the
strength of the inter-species NOL and $B_i(\bar{x})$ is directly
related to the atomic scattering length of the species $i$
($B_i(\bar{x})=2a_{s,i}\hbar\omega_\perp$). In the following we
fix the spatial dependence of $B_{i}(\bar{x})$ and
$S_{12}(\bar{x})$ as
\begin{eqnarray}
B_i &\equiv& B_i(\bar{x})=\Gamma_{i0}+\Gamma_{i}\cos(2k\bar{x})\nonumber\\
S_{12}&\equiv& S_{12}(\bar{x})=G_{i0}+ G_i\cos(2k\bar{x})
..\label{Gammas}\end{eqnarray}

Introducing the dimensionless variables
$$x = \bar{x}k, t = \bar{t}\omega_{R}, \omega_R = \frac{E_R}{\hbar}, E_R = \frac{\hbar^2 k^2}{2m},
u_i = \sqrt{\frac{|\Gamma_{i0}|}{E_R}}\psi_{i},$$ and
\begin{eqnarray}
\beta_i &\equiv& \beta_i(x)=\gamma_{i0}+\gamma_{i}\cos(2x)\nonumber\\
\sigma_{12}&\equiv& \sigma_{12}(x)=g_{i0}+g_{i}\cos(2x)
,\label{gammas}\end{eqnarray}

where $\gamma_{i0} = \Gamma_{i0}/|\Gamma_{i0}|$, $\gamma_i =
\Gamma_i/|\Gamma_{i0}|$, $g_{i0}=G_{i0}/|\Gamma_{j0}|$, $g_{i0} =
G_{1}/|\Gamma_{j0}|$. Below we will consider the particular case
$g_{10} = g_{20} = g_0$, $g_{1}=g_2$.

We can rewrite the above pair of equations as

\begin{eqnarray} \label{GP}
{\rm i}\frac{\partial }{\partial t} \left(\begin{array}{c} {u_1}\\
{u_2} \end{array} \right) &=& - \frac{\partial^2 }{\partial x^2}
\left(\begin{array}{c}u_1\\ u_2 \end{array} \right) \nonumber\\
&+& \left(\begin{array}{cc}
\beta_1|u_1|^2 & \sigma_{12}u_2^* u_1\\
\sigma_{12}u_1^* u_2 &\beta_2|u_2|^2
\end{array}\right)
\left(\begin{array}{c}u_1\\ u_2 \end{array} \right)
,
\end{eqnarray}
where the normalization of the total wave-function $\Psi$  is
related to the components $u_i$ and to the number of atoms $N_i$
by the equation
\begin{equation}
\int_{-\infty}^{\infty}\Psi^\dagger\Psi dx =
\int_{-\infty}^{\infty}dx \left( u_1^* \;\; u_2^* \right)
\left(\begin{array}{c} {u_1}\\ {u_2} \end{array} \right) = N_1 +
N_2. \label{norm}\
\end{equation}

We remark that in experiments the magnitude and sign of both the
inter- and  the intra-species scattering lengths can be controlled
by external magnetic fields \cite{Inouye} or by counter
propagating laser fields \cite{Simoni,OFR1}. In the case of
immiscibility, when $g_{12} < \sqrt{|g_{11}g_{22}|}$, the
repulsive cross-interaction between the components affects
strongly the self-interaction between components.
\subsection{Variational Approach}
In this section we perform an analytical study in the framework of
the variational approach (VA) for the  case of localized (soliton)
solutions of the form: $u_i(x,t)= u_i(x)\exp(-{\rm i}\mu_i t)$,
where $\mu_i$ are the chemical potentials. From Eq.~(\ref{GP}) we
have
\begin{eqnarray}
\label{GPE}
\mu_i u_i &=& - \frac{\partial^2 u_i}{\partial x^2} +
(\gamma_{i0} + \gamma_{i}\cos(2x))u_i^3 +
\nonumber\\ &+&
(g_0 + g_1 \cos(2x) ) u_{3-i}^2 u_i .
\end{eqnarray}
The total energy can be obtained from Eqs.~(\ref{GP}) and (\ref{norm}):
\begin{eqnarray}\label{energy0}
E&=&\frac{\langle \Psi\left|H\right|\Psi\rangle}{\langle\Psi|\Psi\rangle}=
\frac{
\int_{-\infty}^{\infty}dx
\left( u_1\;\; u_2 \right) H
\left(\begin{array}{c}
{u_1}\\ {u_2} \end{array} \right)
}
{N_1 + N_2},\end{eqnarray}
where
\begin{eqnarray}
H&=&
\left(\begin{array}{cc}
-\frac{\partial^2 }{\partial x^2}+\frac{\beta_1}{2}|u_1|^2 &
\frac{\sigma_{12}}{2}u_2^*u_1\\
\frac{\sigma_{12}}{2}u_1^* u_2 &-\frac{\partial^2 }
{\partial x^2}+\frac{\beta_2}{2}|u_2|^2
\end{array}\right),\label{H}
\end{eqnarray}

\begin{eqnarray}\label{energy}
E
&=&\left\{
\int_{-\infty}^{\infty}dx
\sum_{i=1}^{2}\left[\left|\frac{\partial u_i}{\partial x}\right|^2
+
\frac{\beta_i(x)|u_i|^4}{2}\right] +\right. \nonumber\\
&+& \left. \int_{-\infty}^{\infty}dx\;
\sigma_{12}(x)\;|u_1|^2|u_2|^2 \right\}\frac{1} {N_1 +
N_2}.
\end{eqnarray}

The corresponding Lagrangian is given by
\begin{eqnarray}\label{lagr}
{\cal L} &=& \sum_{i=1}^2 \left[ \frac{{\rm i}\hbar}{2}\left(
u_i^*\frac{\partial u_i}{\partial t} -
u_i \frac{\partial u_i^*}{\partial t}
\right)
- \left|\frac{\partial u_i}{\partial x}\right|^2\right.
\nonumber\\
&-& \left. \frac{\beta_i(x)|u_i|^4}{2}\right]
-\sigma_{12}(x)|u_1|^2 |u_2|^2 \nonumber \\
&=& \sum_{i=1}^2 \left[ \mu_i |u_i|^2
- \left|\frac{\partial u_i}{\partial x}\right|^2
-\frac{
(\gamma_{i0}+\gamma_i\cos(2x))
|u_i|^4}{2}\right]
\nonumber\\ &-&
(g_0+g_1\cos(2x))|u_1|^2 |u_2|^2 .
\end{eqnarray}
In our variational approach we consider $u_i$ given by
\begin{equation}
u_i = \sqrt{\frac{N_i}{\sqrt{\pi} a_i}}\exp\left(-\frac{
[x + (3/2-i) {x_0}]^2}{2a_i^2}\right)\;\;(i=1,2),
\end{equation}
where the normalization $N_i$ is related to the number of atoms of 
the species $i$, $a_i$ is the corresponding width, and $x_0$ is a 
parameter given the relative initial position of the two components.
By substituting this ansatz in Eq.~(\ref{energy}) and in the
averaged Lagrangian $L = \int_{-\infty}^{\infty} {\cal L}dx$, we
obtain:
\begin{eqnarray}\label{energy1}
E &=& \left\{
\sum_i \left[\frac{N_i}{2a_i^2} + \frac{N_i^2}{\sqrt{\pi}}\Gamma_i\right]
+\frac{N_1N_2}{\sqrt{\pi}}G\right\}\frac{1}{N_1+N_2} ,
\end{eqnarray}

\begin{eqnarray}
L &=&\sum_i
\left[\mu_i N_i - \frac{N_i}{2a_i^2} -
\frac{N_i^2}{\sqrt{\pi}}\Gamma_i
\right] -
\frac{N_1 N_2}{\sqrt{\pi}}G,
\end{eqnarray}
where
\begin{eqnarray}
\Gamma_i &\equiv& \Gamma_i(a_i,x_0)=
\frac{\gamma_{i0} + \gamma_i e^{-a_i^2/2}\cos(x_0)}{\sqrt{8}a_i},\\
G &\equiv& G(a_1^2,a_2^2, x_0 ) =
\frac{1}{\sqrt{a_1^2 + a_2^2}}e^{-x_0^2/(a_1^2+ a_2^2)}\\
&\times& \left[g_0 + g_1 e^{-\frac{a_1^2 a_2^2}{a_1^2 +
a_2^2}}\cos(x_0 \frac{a_2^2 - a_1^2}{a_1^2 +
a_2^2})\right].\nonumber
\end{eqnarray}
 From the Euler-Lagrange equations $\partial L/\partial N =0$,
 $\partial L/\partial a =0$ and $\partial L/\partial x_0 =0 $ we obtain
 the equations for the chemical potentials $\mu_i$ and number
 of atoms $N_i$:
\begin{eqnarray}\label{mui}
\mu_i &=& \frac{1}{2a_i^2} + \frac{N_i}{\sqrt{\pi}}2\Gamma_i
+\frac{N_j}{\sqrt{\pi}}G,\\
\frac{N_i}{\sqrt{\pi}} &=& \left[\frac{P_{3-i}-Q_i}{P_1P_2-Q_1Q_2}\right],
\end{eqnarray}
with
\begin{eqnarray}
P_i&\equiv& {a_i^3}\frac{\partial \Gamma_i}{\partial a_i},\;\;\;
Q_i\equiv {a_i^3}\frac{\partial G}{\partial a_i}\\
Q_i&=&{2a_i^4}\left\{
\frac{\left(x_0^2-\frac{a_1^2+a_2^2}{2}\right)}{(a_1^2+a_2^2)^2}G
+
\frac{\left(2x_0\tan(x_0\frac{a_2^2-a_1^2}{a_1^2+a_2^2})-a_j^4\right)}
{(a_1^2+a_2^2)^{5/2}}\times \right. \nonumber\\&& \left. g_1\left(
e^{-\frac{x_0^2+a_1^2 a_2^2}{a_1^2 + a_2^2}}\cos(x_0 \frac{a_2^2 -
a_1^2}{a_1^2 + a_2^2}) \right)\right\},
\\
P_i &\equiv& -\frac{a_i}{\sqrt{8}}
\left[\gamma_{i0} + (1 + a_i^2)\gamma_i e^{-a_i^2/2}\cos(x_0)\right],
\end{eqnarray}
\begin{eqnarray}
\sin(x_0) &=& \frac{\sqrt{8}N_1 N_2 (\partial{G}/\partial{x_0})}
{N_1^2\gamma_1 (e^{-a_1^2/2}/a_1) + N_2^2\gamma_2 (e^{-a_2^2/2}/a_2)},
\end{eqnarray}
where $i \neq j = 1,2 .$
From (\ref{mui}) and (\ref{energy1}), it also follows that
\begin{equation}\label{en2}
E=\frac{1}{2(N_1+N_2)}\sum_iN_i\left(\mu_i+\frac{1}{2a_i^2}\right).
\end{equation}

For the particular choice of parameters for the symmetric case,
when $\gamma_{10} = \gamma_{20}=\gamma_0$, $\gamma_1 = \gamma_2 =
\gamma$, we have $\mu_i=\mu_j=\mu$, $a_i=a_j=a$, $N_i=N_j=N$, and
$u_{i=1,2}(x,t)\equiv u_\pm(x,t)$, with
\begin{equation}
u_\pm(x,t)=
\sqrt{\frac{N}{a\sqrt{\pi}}}\exp\left(-\frac{(x\pm x_0/2)^2}{2a^2}
\right)
e^{-{\rm i}\mu t}.
\end{equation}
 The equations for the chemical potential $\mu$, energy $E$,  and 
 number of atoms $N$ become
($x_0 \neq 0$)
\begin{eqnarray}
\mu &=& \frac{1}{2a^2} + \frac{N}{\sqrt{2\pi}\;a}
\left[\gamma_{0}+\gamma\cos(x_0)e^{-a^2/2}\right] +
\nonumber \\&+& \frac{N}{\sqrt{2\pi}\;a}
\left[(g_0 + g_1e^{-a^2/2})e^{-x_0^2/2a^2}\right],\label{mus}
\\
E &=& \frac{\mu}{2}+\frac{1}{4a^2},
\\
\frac{N}{\sqrt{2\pi} a}&=&\frac{1}{\sqrt{2}a(P+Q)},\;\;\;\;\;
(Q\equiv Q_i, P\equiv P_i)
\label{Ns}
\end{eqnarray}
\begin{eqnarray}
&\displaystyle\frac{N}{\sqrt{2\pi} a}&=
\frac{-1}{2}\left\{a^2
\left(\gamma_0+\gamma\cos(x_0)e^{-a^2/2}(1+a^2)\right)+
\right.\nonumber\\ &+&\left.
(a^2-x_0^2)\left[g_0+g_1 e^{-\frac{a^2}{2}}\left(1+\frac{a^4}{a^2-x_0^2}\right)
\right] {e^{-\frac{x_0^2}{2a^2}}}
\right\}^{-1}
\nonumber
\end{eqnarray}
\begin{eqnarray}
\sin(x_0)  &=& \frac{\sqrt{2}\;a}{\gamma} e^{(a^2/2)}
\frac{\partial{G}}{\partial{x_0}} \nonumber\\
&=& -\frac{x_0}{\gamma a^2}
\left[\left(g_0e^{a^2/2}+g_1\right)
{\exp\left(-\frac{x_0^2}{2a^2}\right)}\right].
\label{Gx0}
\end{eqnarray}
For $x_0=0$, we have
\begin{eqnarray}
\mu &=& \frac{1}{2a^2} +
\frac{N}{\sqrt{2\pi}\;a}
\left[\gamma_{0}+g_0 + (\gamma+g_1)e^{-a^2/2}\right],
\label{mu1}
\end{eqnarray}
\begin{eqnarray}
\frac{N}{\sqrt{2\pi} a}&=&
\frac{-2}{a^2
\left[\gamma_0+g_0+(\gamma+g_1) e^{-a^2/2}(1+a^2)\right]}
\label{N1} .
\end{eqnarray}
By using (\ref{N1}) in (\ref{mu1}) for the
symmetric case with $x_0=0$ we obtain
\begin{eqnarray}
\mu &=& \frac{-3}{2a^2}
\left[
\frac
{(\gamma_{0}+g_0) + (\gamma+g_1)e^{-a^2/2}(1-\frac{a^2}{3})}
{(\gamma_{0}+g_0) + (\gamma+g_1)e^{-a^2/2}(1+a^2)}\right]
\label{mu2}.
\end{eqnarray}
%
\begin{figure}
\centerline{
\includegraphics[width=7.5cm,height=5.5cm,clip]{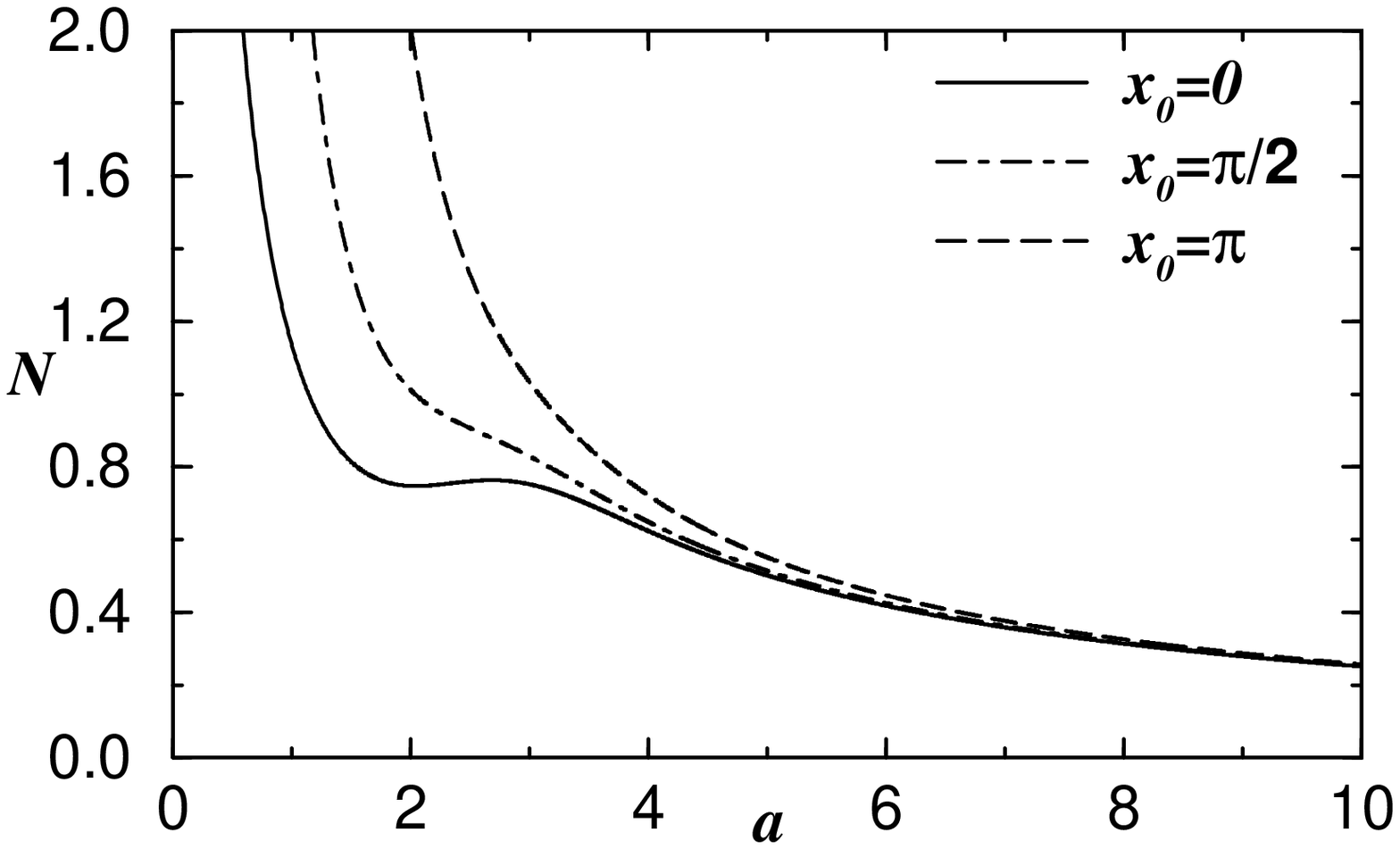}}
\centerline{\includegraphics[width=8cm,height=5.5cm,clip]{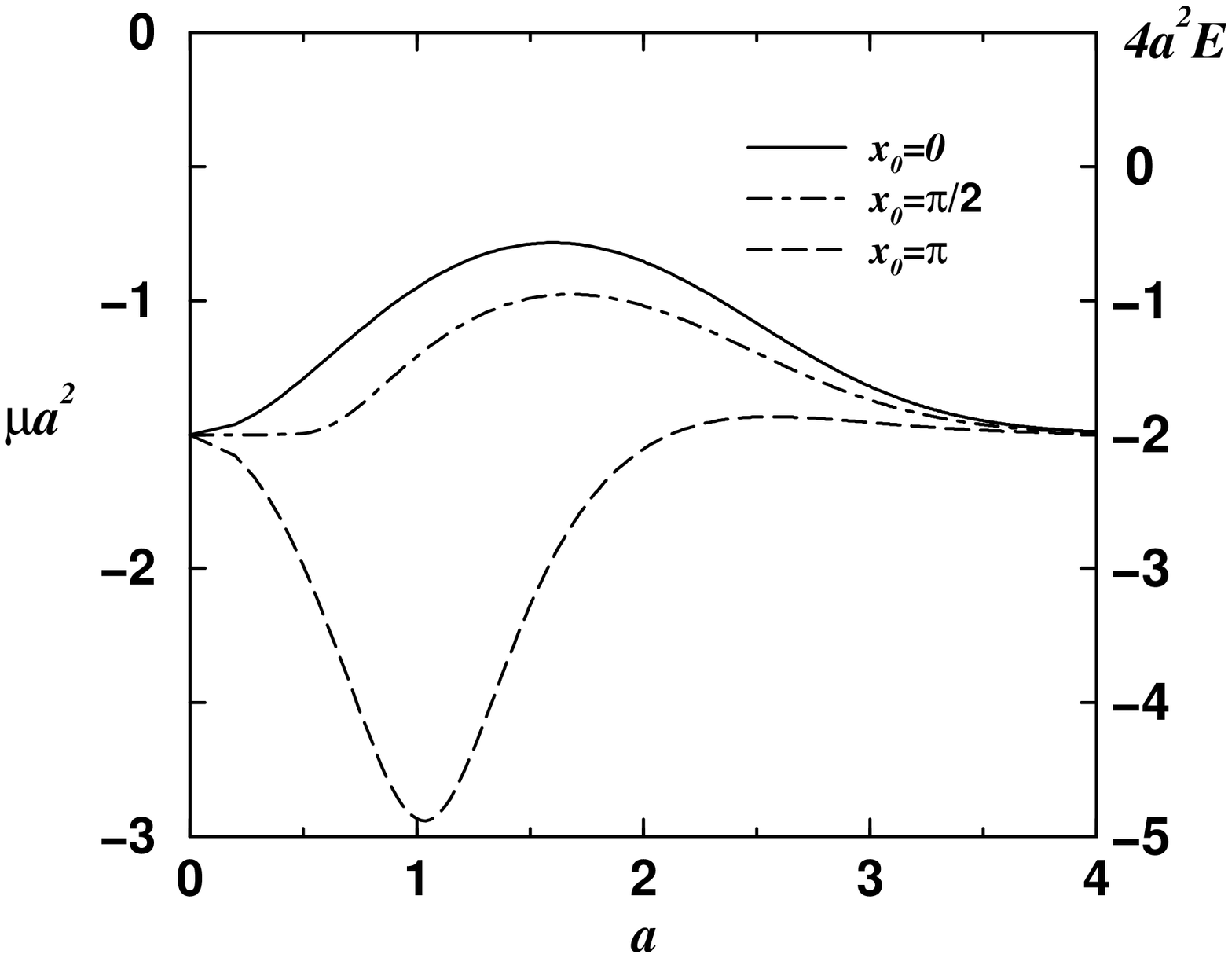}
} \caption{
Considering the symmetric case with $\gamma_{10}=\gamma_{20}=-1$, $\gamma_1=\gamma_2=-0.5$, $g_0=-1$, $g_1=-1.5$,
with different values of $x_0$, we show the 
VA results for the number of particles $N$, chemical potential
$\mu$ and energy $E$, versus the width $a$. $N$ 
is given in the upper panel, with $\mu a^2$ (scale in the lhs) and 
$4 a^2 E$ (scale in the rhs) given in the lower panel. 
} 
\label{Fig01}
\end{figure}
%
\begin{figure}
\centerline{
\includegraphics[width=7.5cm,height=5.5cm,clip]{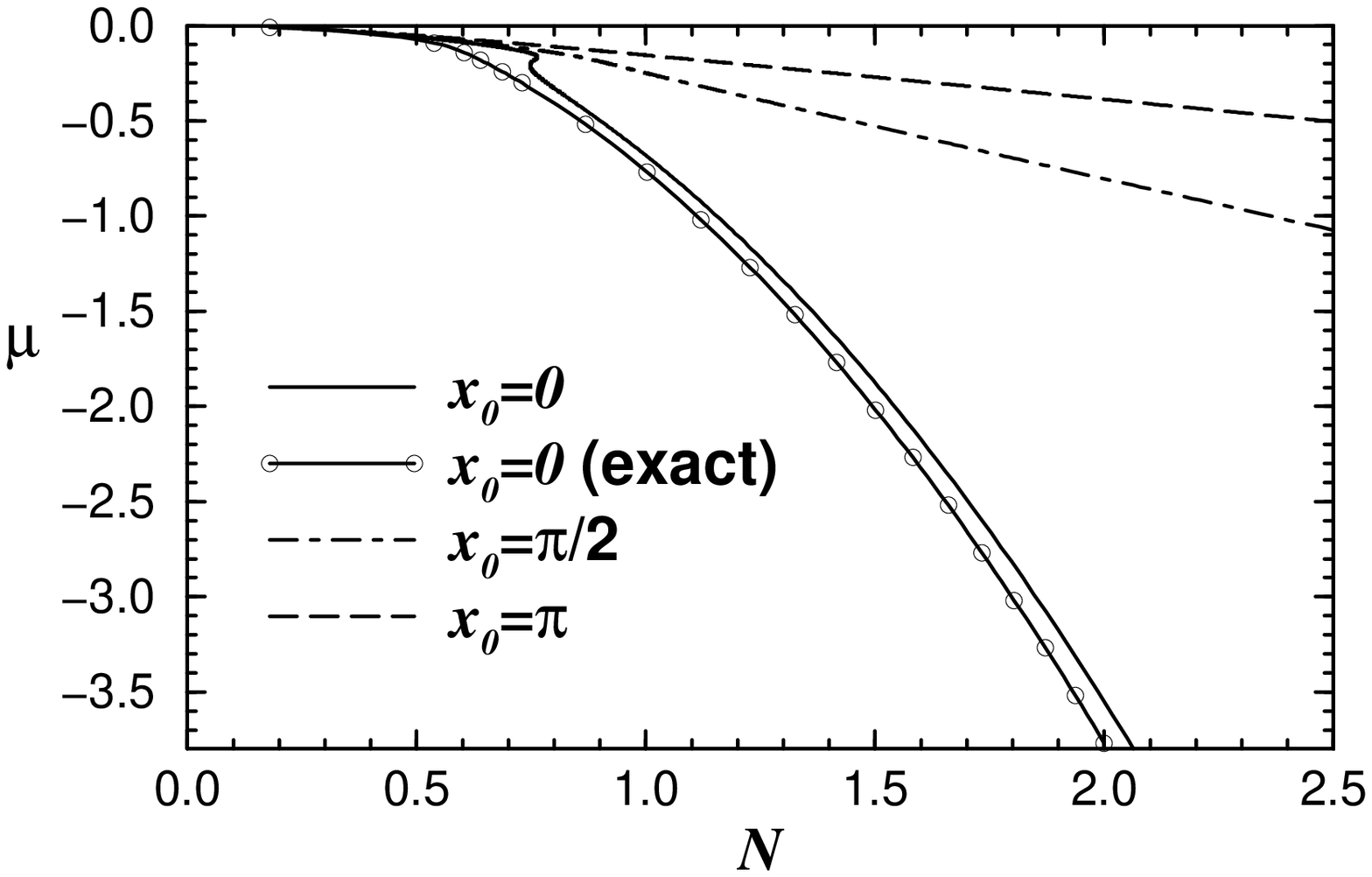}}
\centerline{\includegraphics[width=7.5cm,height=5.5cm,clip]{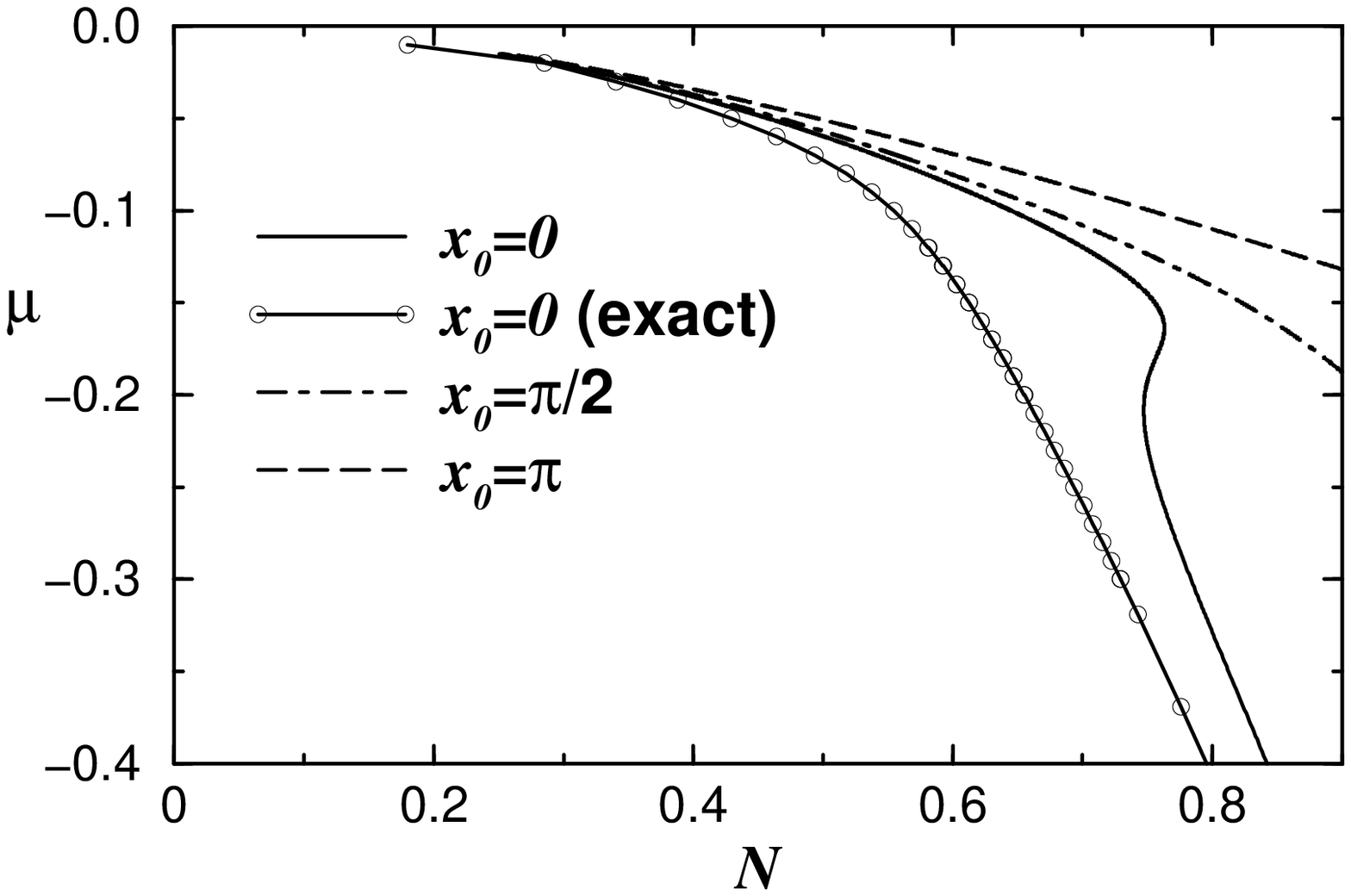}
} \caption{VA solution for the chemical potential versus the
number of particles $N$, in the symmetric case with the same
parameters as in Fig.~\ref{Fig01}. Exact results are also
shown for the case of $x_0=0$ (solid line with empty circles, in both panels). 
Here we observe that the small unstable region, $d\mu/dN > 0$, presented by 
the VA is not confirmed by the full numerical results.} 
\label{Fig02}
\end{figure}
%
\begin{figure}
\centerline{
\includegraphics[width=7.3cm,height=5.5cm,clip]{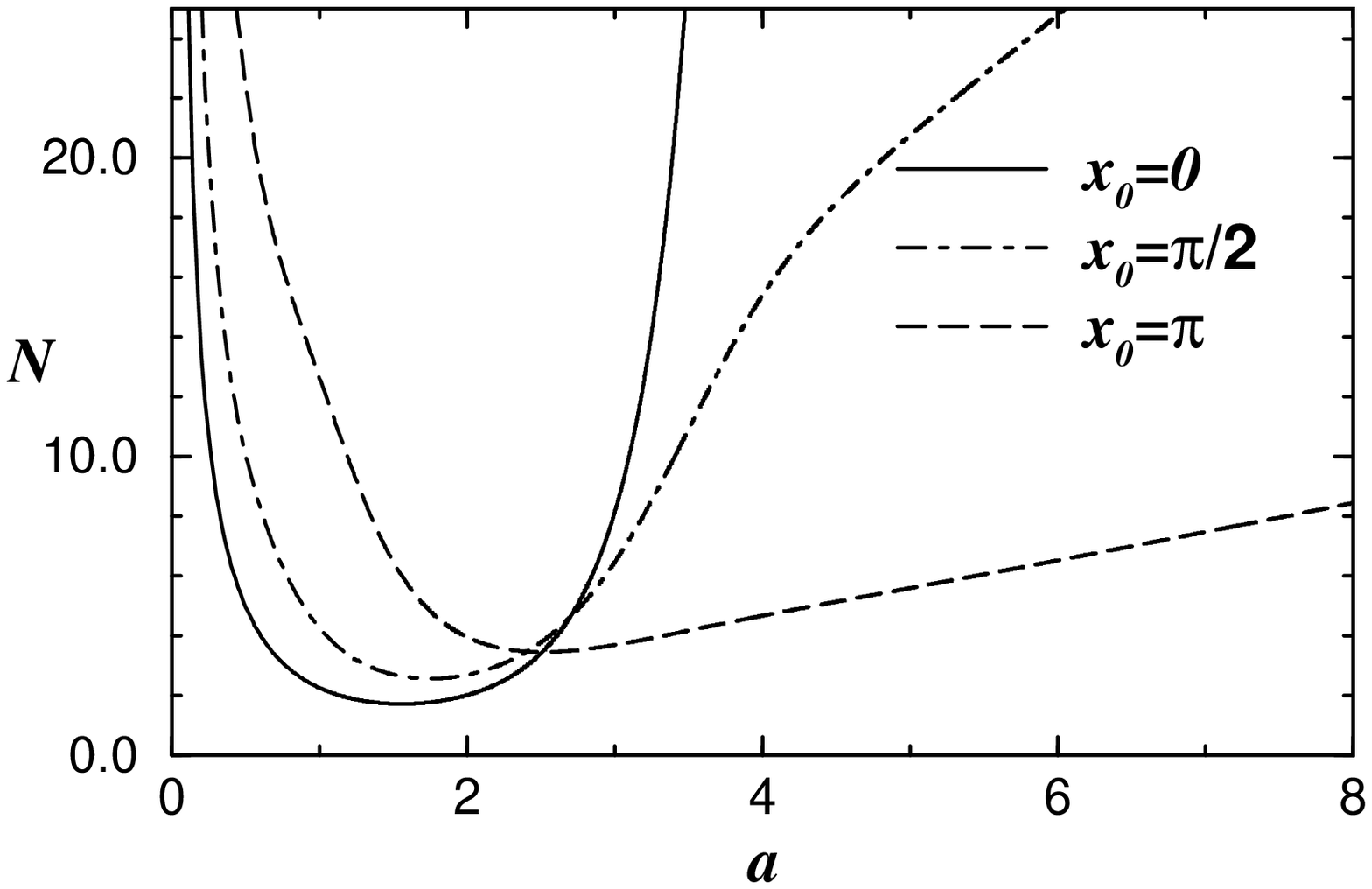}}
\centerline{\includegraphics[width=7.5cm,height=5.5cm,clip]{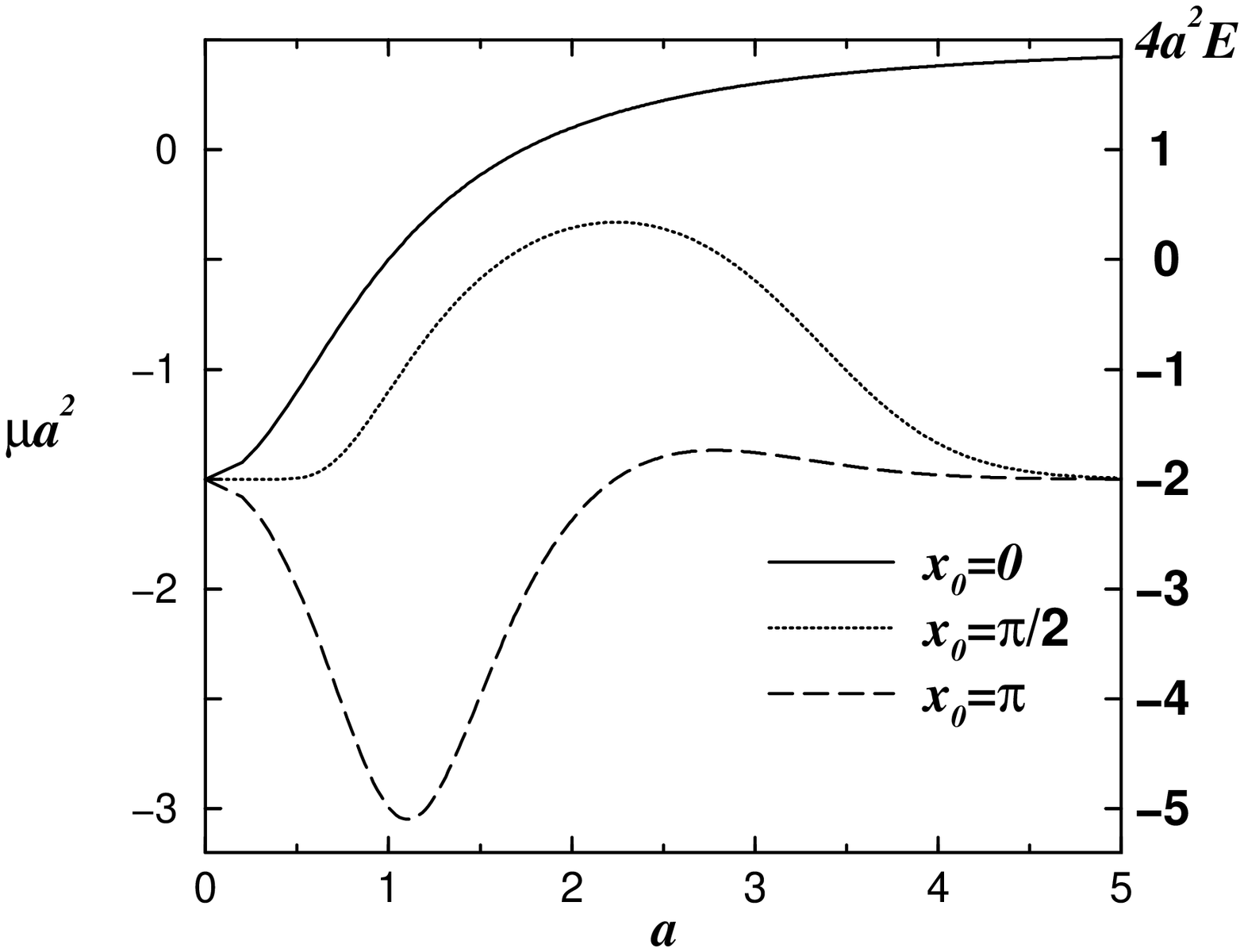}
} \caption{
Considering the symmetric case with $\gamma_{10}=\gamma_{20}=-1$,
$\gamma_1=\gamma_2=-0.5$, $g_0=1$, and $g_1=-1.338926$,
with different values of $x_0$, we show the 
VA results for the number of particles $N$, chemical potential
$\mu$ and energy $E$, versus the width $a$. $N$ 
is given in the upper panel, with $\mu a^2$ (scale in the lhs) and 
$4 a^2 E$ (scale in the rhs) given in the lower panel. 
}\label{Fig03}
\end{figure}
%
\begin{figure}
\centerline{
\includegraphics[width=7.5cm,height=5.5cm,clip]{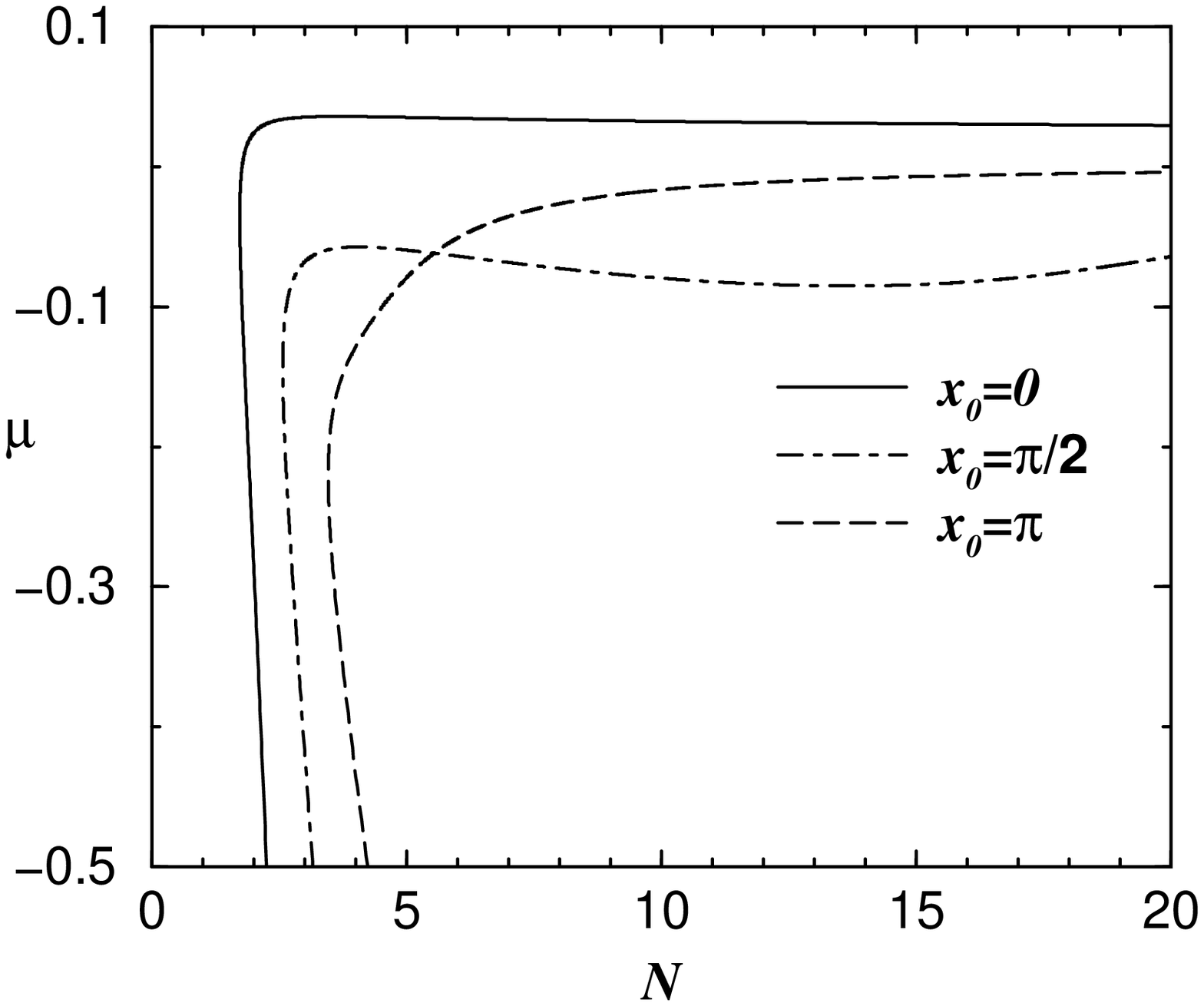}}
\centerline{\includegraphics[width=7.5cm,height=5.3cm,clip]{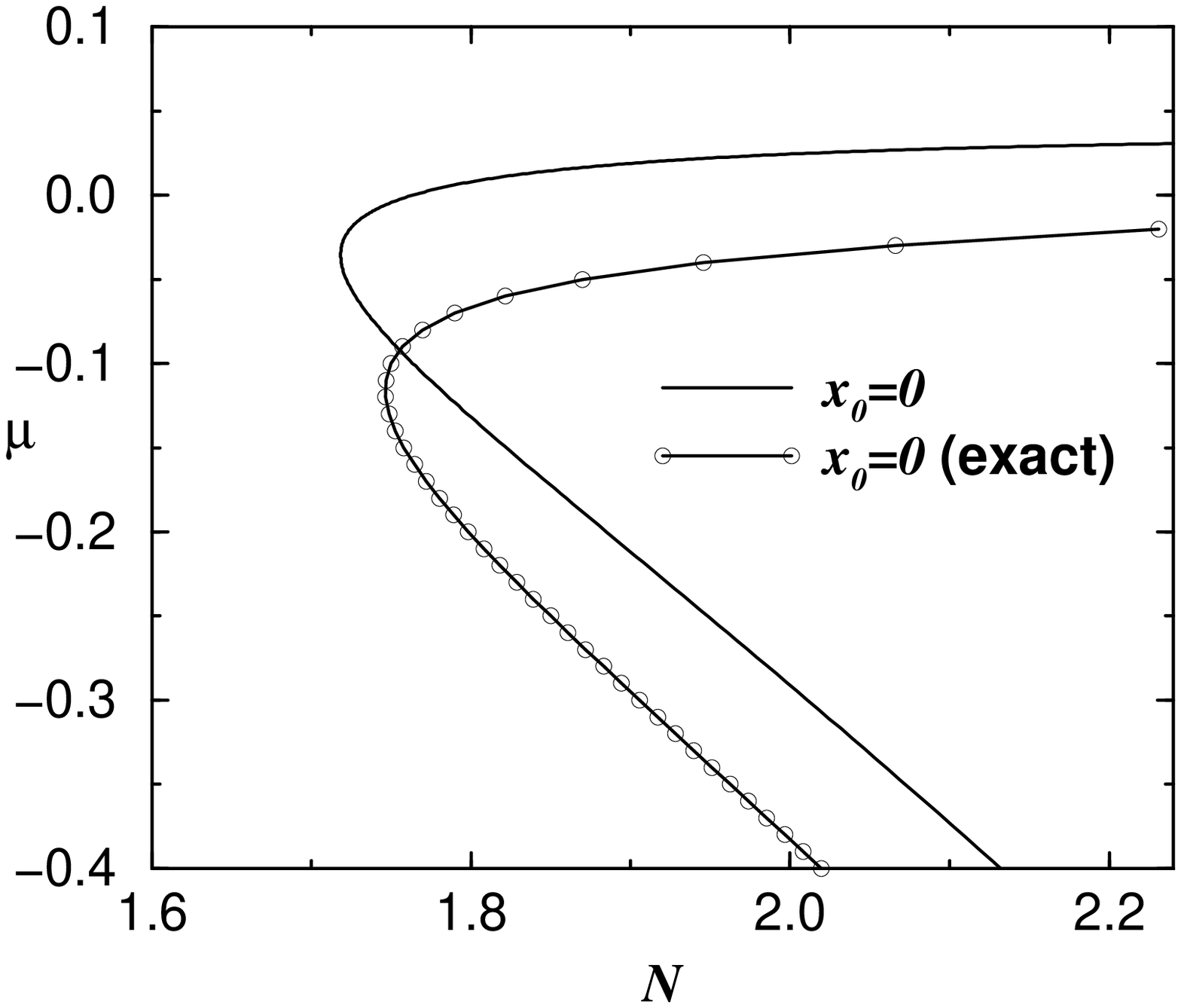}}
\caption{VA solution for the chemical potential versus the number
of particles $N$, in the symmetric case with the same parameters
as in \ref{Fig03}. As we can see, the stable region ($d\mu/dN
> 0$) is more pronounced for $x_0=\pi$ than for $x_0=0$.}
\label{Fig04}
\end{figure}

The stability of the soliton solution can be investigated by using
the Vakhitov-Kolokolov criterion~\cite{VK} (in the present case,
implying that for a stable system we should have $dN/d\mu <0$),  and
also by studying the total energy $E$ and chemical potentials $\mu_i$ 
as functions of the width $a$. For the symmetric cases (when $N_1=N_2=N$
and $\mu_1=\mu_2=\mu$), the results of such study is presented in 
Figs.~\ref{Fig01} to \ref{Fig04}, considering an attractive inter-species
scattering length ($g_0<0$) in Figs.~\ref{Fig01} and \ref{Fig02}; and
repulsive inter-species scattering length ($g_0>0$) in Figs.~\ref{Fig03} 
and \ref{Fig04}.
From Figs.~\ref{Fig01} and \ref{Fig03} we obtain the behavior of $N$, 
chemical potential $\mu$ and energy $E$ as functions of $a$.
The behavior of $\mu$ versus $N$, in order to check the VK criterion, 
is shown in Figs. \ref{Fig02} and \ref{Fig04}.
This stability study was done mainly by using the variational approach (VA),
considering different values of the parameter $x_0$, which gives the position
of the soliton solution in respect to the optical lattice.
In case of $x_0$ the VA solutions are also compared with full numerical results
in Figs. \ref{Fig02} and \ref{Fig04} (solid lines with empty circles). 
As observed, the VA gives a good qualitative picture of the exact results, 
with improved quantitative results for large values of $|\mu|$.

The dominant $1/a^2$ behavior of the chemical potential $\mu$ 
and energy $E$,  as functions of the width $a$,  are removed in the 
bottom panels of Figs.~\ref{Fig01} and \ref{Fig03} (by a multiplicative  
factor proportional to $a^2$), in order to enhance their $x_0$ dependence.  
As we can verify, in both the cases, the most stable configuration is 
obtained when $x_0=\pi$.
 
As we can see in Fig.~\ref{Fig02}, the single soliton is stable for
$\gamma_{10}=\gamma_{20} =-1, \gamma_{1}=\gamma_2 = -0.5, g_0 =
-1, g_1 =-1.5$. The VA predicts the existence of small instability
region, that is not confirmed by the numerical simulations of the
system of GP equations. This instability region corresponds to the
broad soliton case with $a/\pi > 1$, where the VA approach is not
applicable. 

\begin{figure}
\centerline{
\includegraphics[width=7.5cm,height=5.5cm,clip]{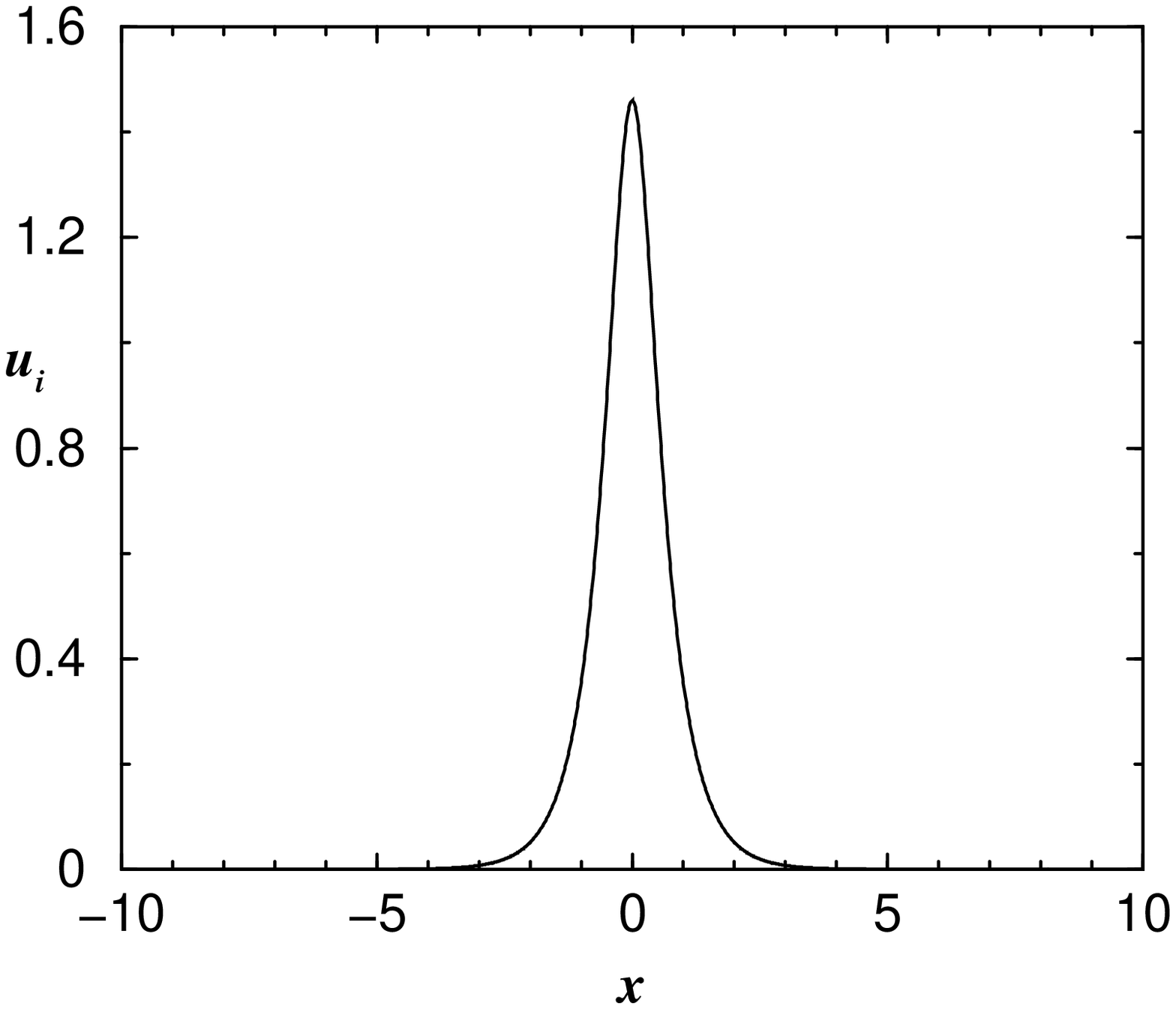}}
\centerline{\includegraphics[width=7.5cm,height=5.5cm,clip]{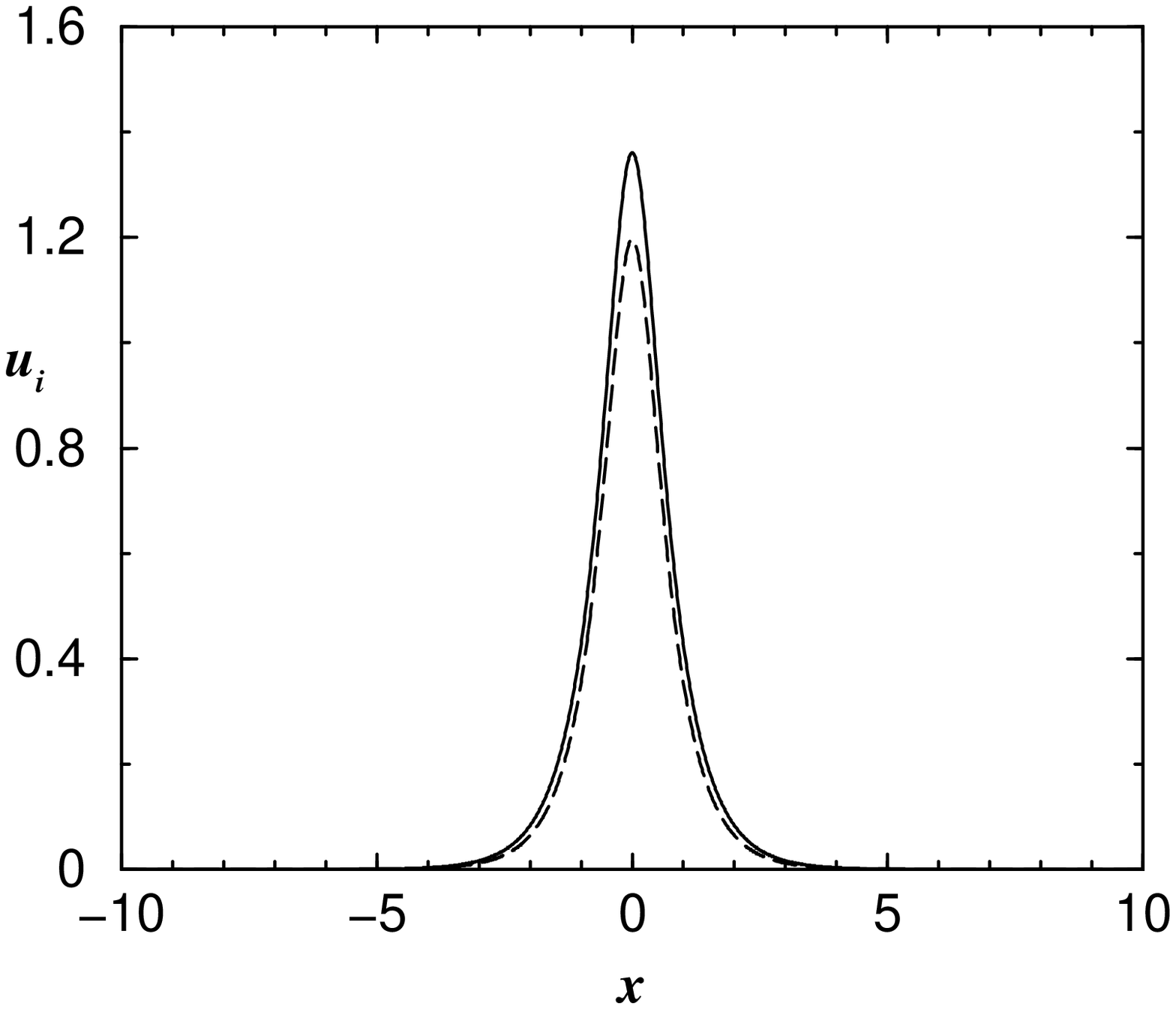}
} \caption{On-site symmetric modes of Eq.~(\ref{GP}) with NOL
parameter values $\gamma_{10}=\gamma_{20}=-1$,
$\gamma_1=\gamma_2=-0.5$, $g_0=-1$, $g_1=-1.5$. The top panel
refers to the case of an equal number of atoms $N_1=N_2=2$ in the
two components with equal chemical potentials
$\mu_1=\mu_2=-3.769$. The bottom panel refers to the case of
different  number of atoms $N_1=2, N_2=1.5$ in the two components
and different  chemical potentials $\mu_1=-2.698, \mu_2=-2.936$.
Dashed lines refer to second components.} 
\label{Fig05}
\end{figure}

\section{Symmetry properties of localized modes}

In this section we investigate the symmetry properties of
localized modes with similar number of particles in each
component. These modes can be of equal symmetry or of mixed
symmetry type. In order to find these solutions we use both the
self-consistent exact diagonalization method and the generalized
relaxing method described in the Appendix (these methods provide
identical results for all the cases studied below, with the only
exception of the state in Fig.~\ref{Fig08}, for which the
relaxation method was not effective).

In Fig.~\ref{Fig05} we show the fundamental modes obtained in the
attractive case ($\gamma_{i0}<0, g_0<0$) with equal and different
number of atoms in the two components.
%
\begin{figure}
\centerline{
\includegraphics[width=7.5cm,height=5.cm,angle=0,clip]{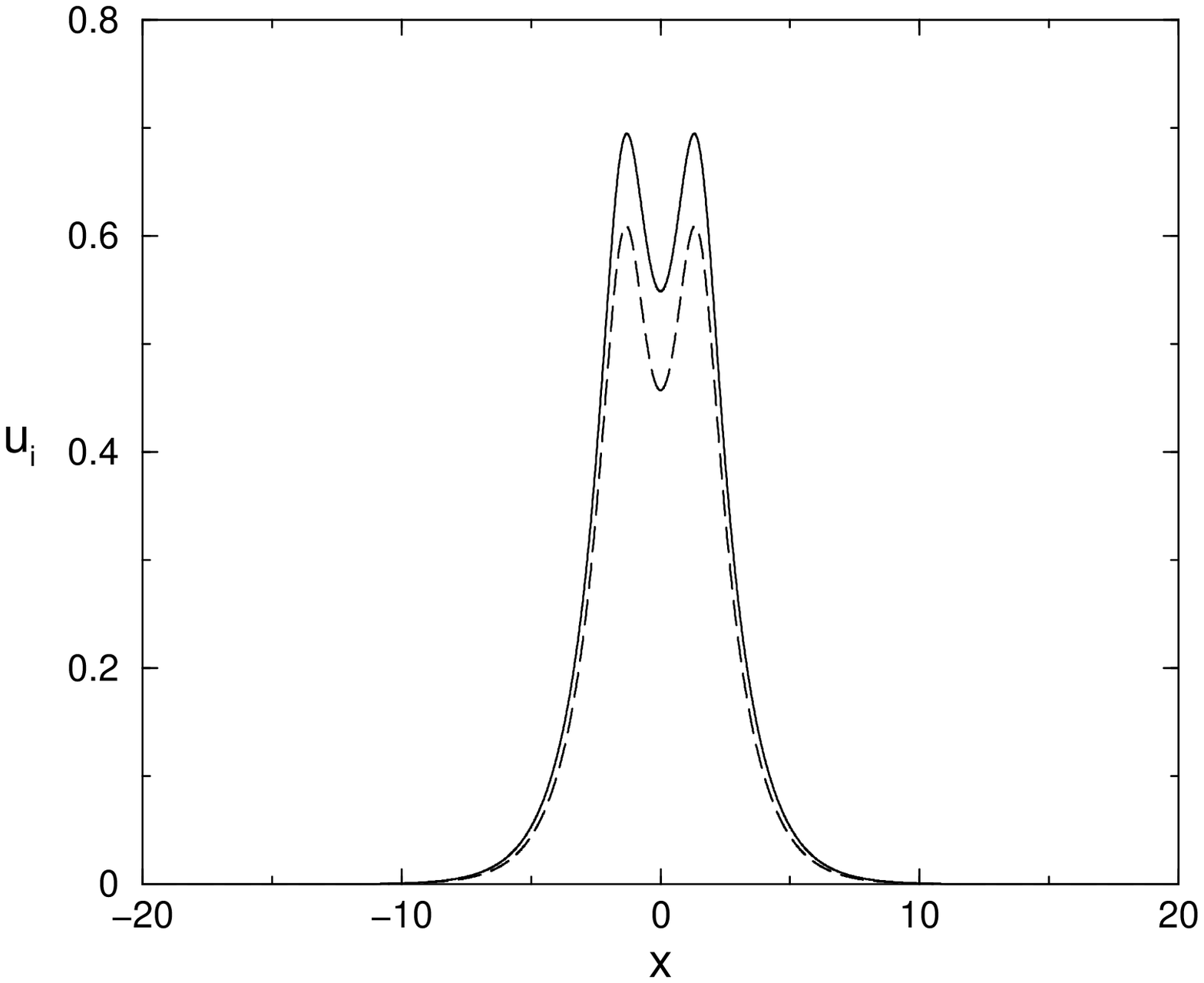}}
\centerline{\includegraphics[width=7.5cm,height=5.cm,angle=0,clip]{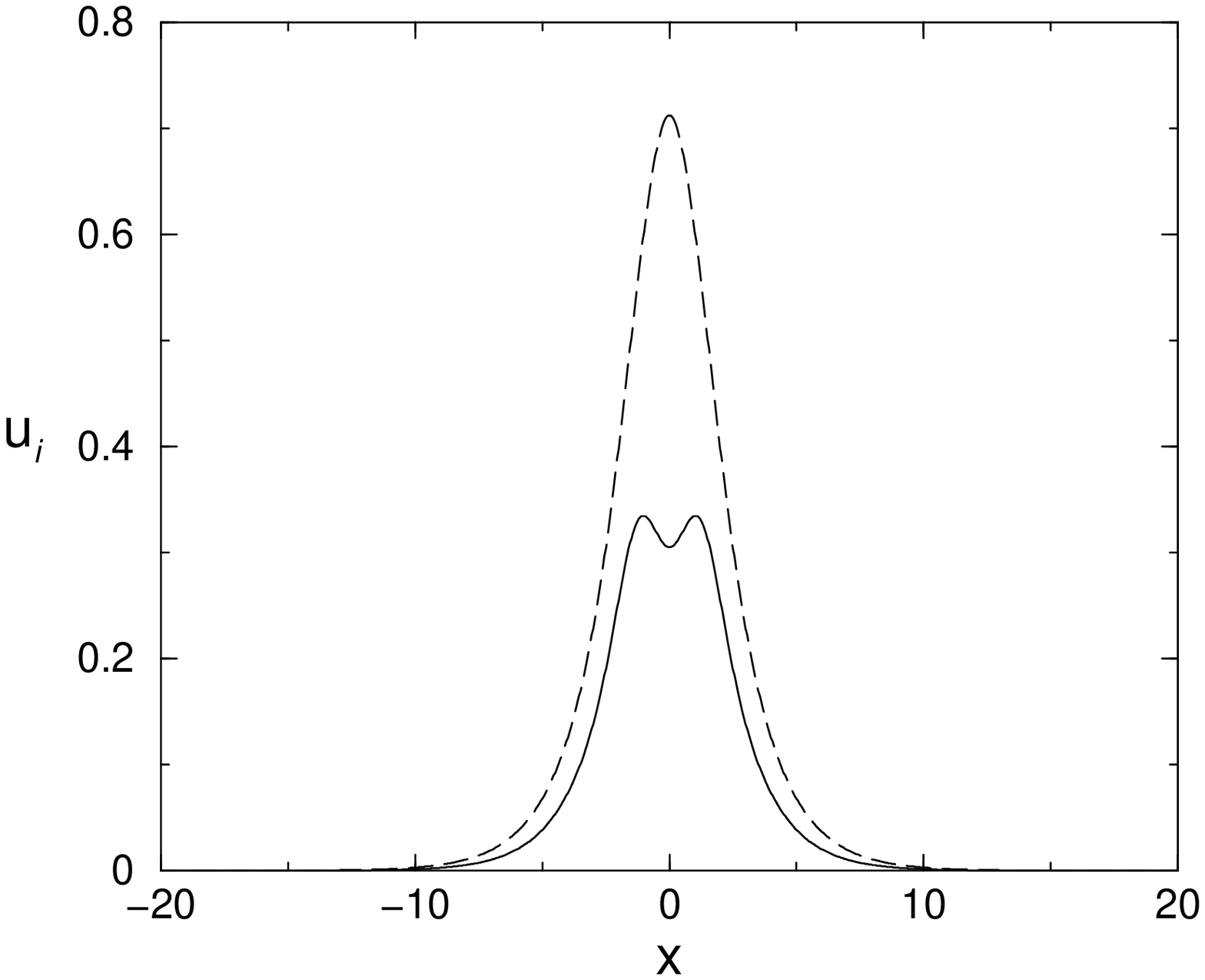}
 } \caption{Localized modes of the two component GPE in Eq.~(\ref{GP}) of
 symmetry type IS-IS (top panel) and IS-OS (bottom
panel). The number of atoms and chemical potentials of the IS-IS
mode are $N_1=2, N_2=1.5$, $\mu_1=-0.653, \mu_2=-0.678 $, while
for the IS-OS mode are $N_1=0.5, N_2=1.7$, $\mu_1=-0.427,
\mu_2=-0.3849$. The parameters of the NOL for the IS-IS mode are
fixed as $\gamma_{10}=\gamma_{20}=-1$, $\gamma_1=\gamma_2=0.5$,
$g_0=-1$, $g_1=1.5$,  while for the IS-OS mode are fixed as:
$\gamma_{10}=\gamma_{20}=-1$, $\gamma_1=0.9, \gamma_2=-0.5$,
$g_0=-1.5$, $g_1=1.5$. Dashed lines refer to second components.}
\label{Fig06}
\end{figure}
%
\begin{figure}
\centerline{
\includegraphics[width=7.5cm,height=5.5cm,angle=0,clip]{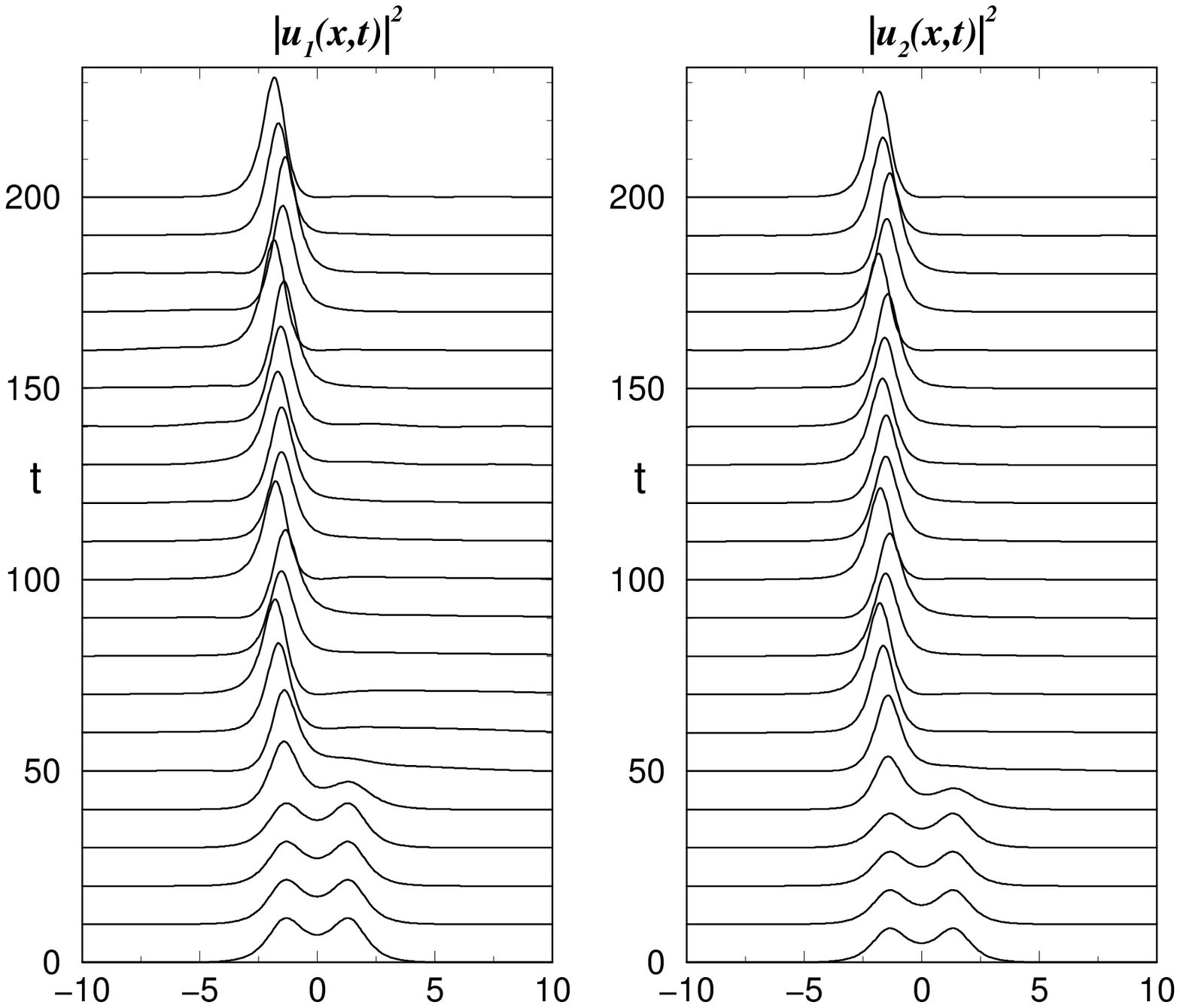}
}\centerline{
\includegraphics[width=7.5cm,height=5.5cm,angle=0,clip]{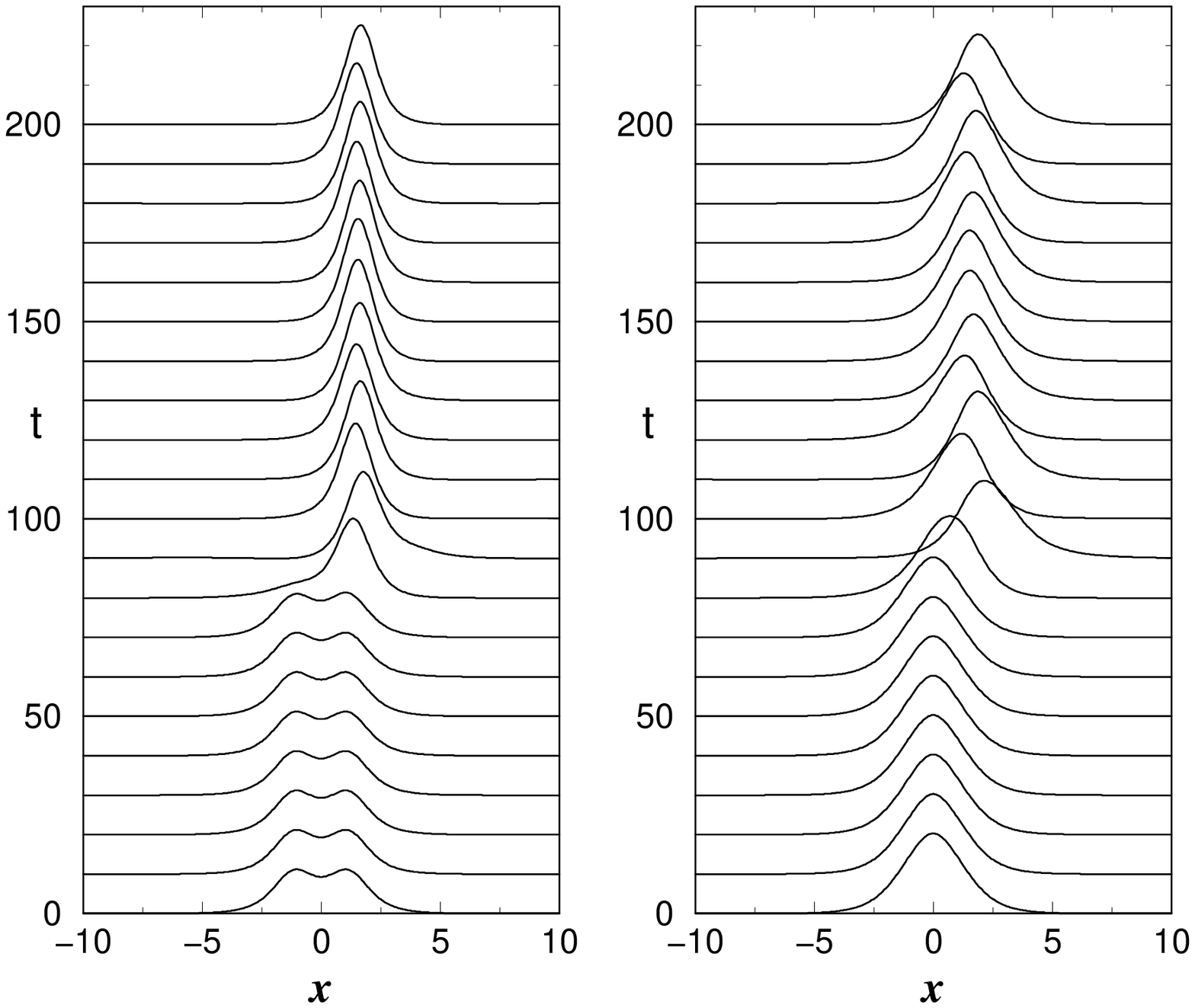}
}
\caption{Time evolution of the modulo square of the IS-IS (top
panels) and OS-IS mode (bottom panels) depicted in
Fig.~\ref{Fig06}. Left panels refer to first components.}
\label{Fig07}
\end{figure}
In both cases we have that the maximum of the atomic densities are
symmetric around the minimum of the corresponding effective
potentials (see Appendix A). Adopting the same terminology
introduced in \cite{Cruz07} for the case of a LOL, we shall refer
to these modes as OS-OS (onsite symmetric in both components).
Note that OS-OS-modes with equal number of atoms  have  the same
chemical potentials, while for different number of atoms the
component with a lower number of atoms has also a lower chemical
potential. For sufficiently strong NOL (see below) these modes are
very stable under GPE time evolution and  represent the
fundamental ground states of the system in the case of  all
attractive interactions. In particular,  the GPE time evolution of
the density of the OS-OS mode  in Fig.~\ref{Fig05} with a
different number of atoms does not show any deviation from the
starting density for a time $t$ going from $0$ to $200.$

Besides onsite symmetry modes, it is also possible to have  modes
that are intersite symmetric (IS) in one or in both components,
i.e. symmetric around a maximum of the effective nonlinear
potential instead than a minimum. Such modes can be of type IS-IS
(intersite symmetric in both components) such as the one shown in
the top panel of Fig.~\ref{Fig06} or of mixed  type (OS-IS or
IS-OS) such as the one shown in the bottom panel of
Fig.~\ref{Fig06}. In contrast with the OS-OS mode,  the intersite
symmetric localized modes are found to be unstable under GPE time
evolution as one can see from  Figs.~\ref{Fig07} and \ref{Fig08}
for IS-IS and OS-IS modes, respectively. Notice that in both cases
the states decay into an OS-OS mode which is the true ground state
of the system, and that in the IS-OS case the decay of the IS
component give rise to internal oscillations (relative motion
between the two final OS components) which can last for a long
time. Internal oscillations of the OS-OS modes can also be excited
trough scattering with other modes.

Besides modes that are localized in both components it is also
possible to couple a localized mode in one component with an
extended mode in the other component  such that the extended state
acts as a periodic potential for the localized mode and forming a
bound state. Such an example is presented in Figs.~\ref{Fig08} and
\ref{Fig09} for the case of a binary mixture with an average repulsive interaction
for the first component ($\gamma_{10}>0$, $|\gamma_{10}|>|\gamma_1|$ ) 
and an average attractive interaction for the second component ($\gamma_{20} <0$, $|\gamma_{20}|>|\gamma_2|$). 
This combination of signs for the interactions makes the ground state 
of the system to be extended for the first component and localized in the 
second one, leading to the formation of the dark-bright bound state depicted 
in the upper panel of Fig.~\ref{Fig08}. 
Another possible solution for the same combination of parameters is
also verified, as shown in the lower panel of Fig.~\ref{Fig08},
with the formation of a bright-bright state, having one bright
solution on top of the background. 
For the considered parameters, both the solutions presented in Fig.~\ref{Fig08}
are quite stable under the GPE time evolution. In Fig.~\ref{Fig09} we show
the time evolution of the dark-bright state shown in the upper panel of
Fig.~\ref{Fig08}.

\begin{figure}
\includegraphics[width=7.cm,height=5.cm,angle=0,clip]{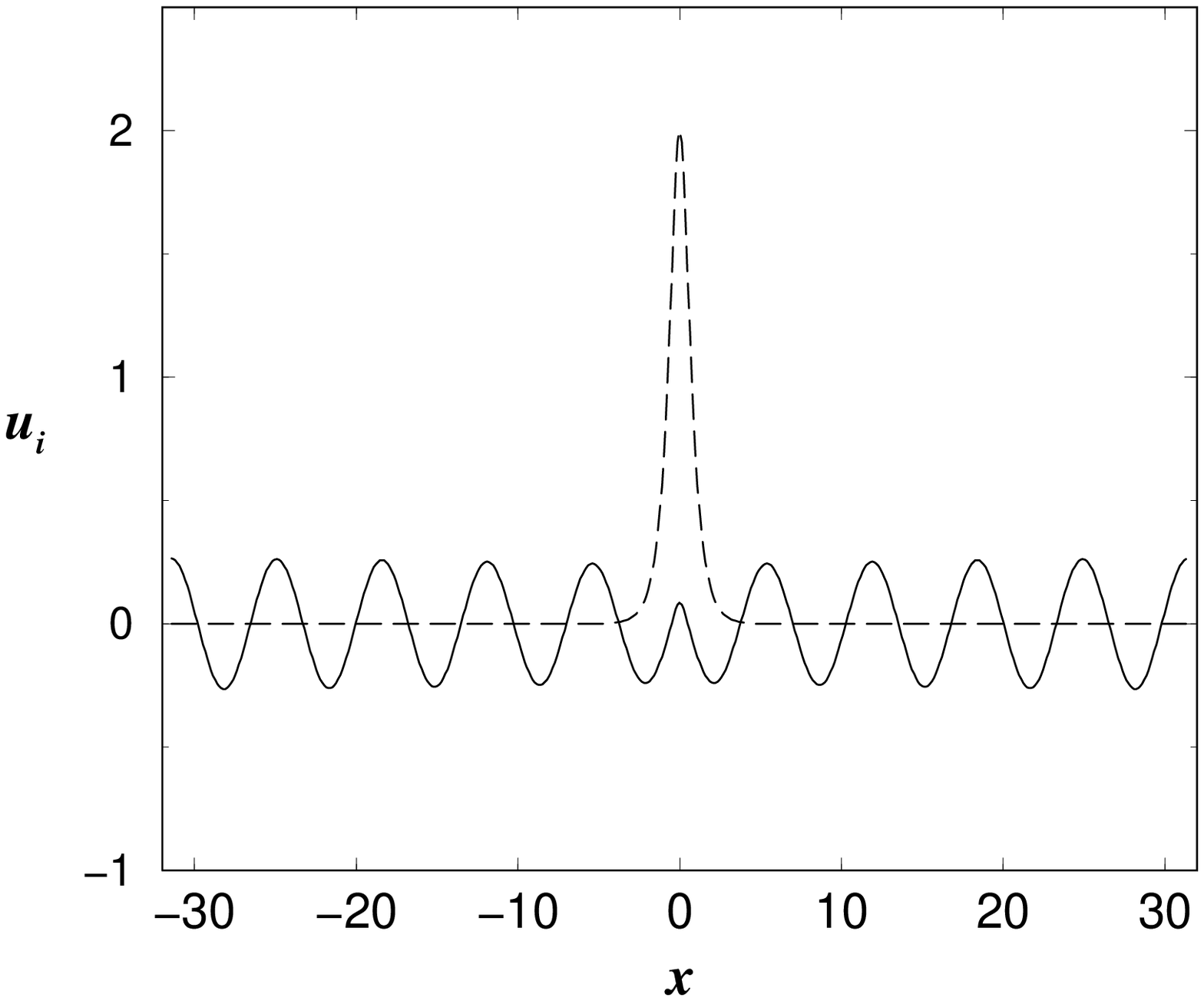}
\includegraphics[width=7.cm,height=5.cm,angle=0,clip]{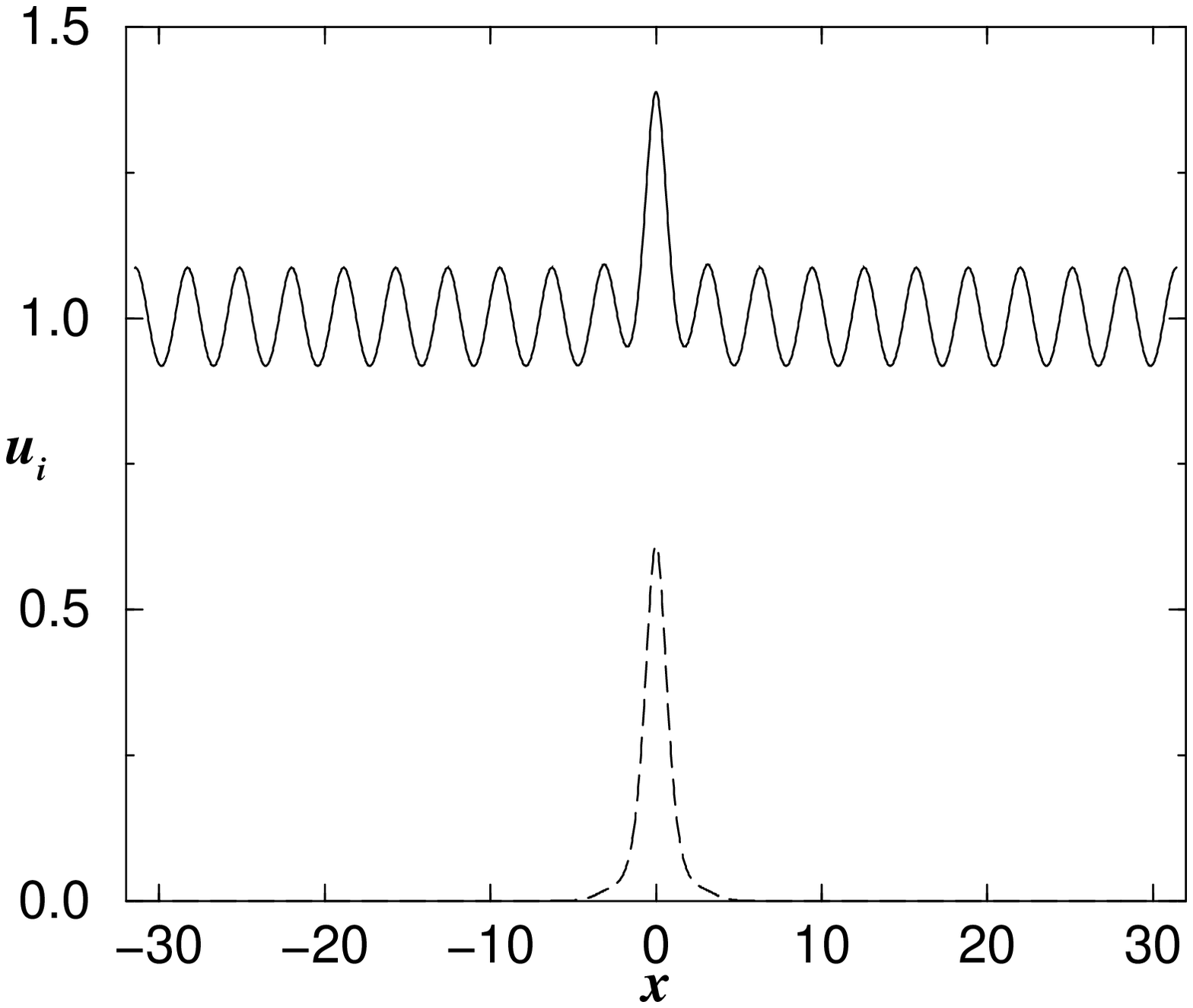}
\caption{Dark-bright (upper panel) and bright-bright (lower panel)
soliton modes of Eq.~(\ref{GP}), obtained with the NOL parameters 
$\gamma_{10}=1, \gamma_{20}=-1$, $\gamma_1=\gamma_2=-0.5$, $g_0=-1$, 
and $g_1=-1.5$. In both the cases, we have the same chemical
potentials $\mu_1=$0.9476 and $\mu_2=-$2.746.
The resulting number of atoms in the two components are,
respectively, $N_1=$2 and $N_2=$4.5 for the dark-bright solution (upper panel);
and $N_1=$64.246 and $N_2=$0.444 for the bright-bright solution (lower panel).}
\label{Fig08}
\end{figure}

\begin{figure}
\centerline{
\includegraphics[width=8.7cm,height=9.7cm,angle=0,clip]{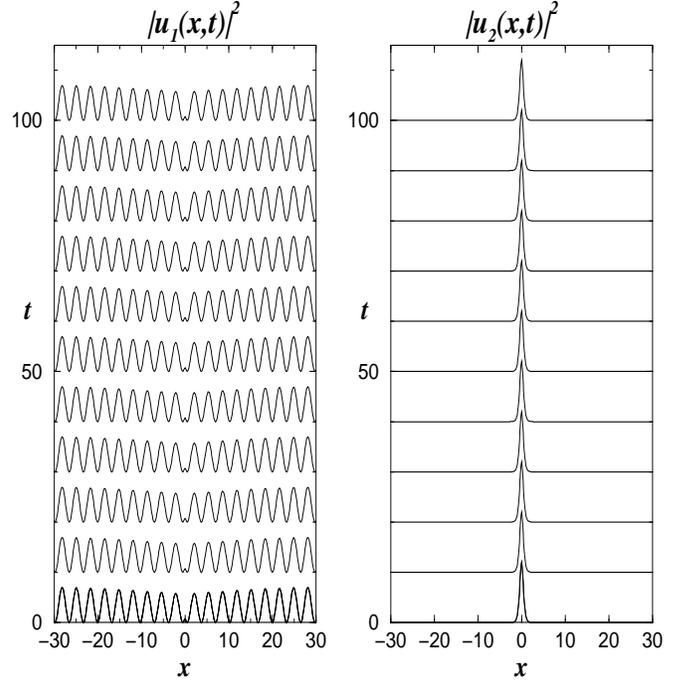}
}
\caption{Time evolution of the dark-bright soliton mode depicted
 in Fig.~\ref{Fig08}.
 } \label{Fig09}
\end{figure}

\section{Delocalizing transition of fundamental OS-OS modes}
In this section we investigate the existence of a delocalizing
transition for the fundamental OS-OS symmetry mode. To this regard
we recall that for a single component 1D BEC with combined LOL and
NOL there exists a threshold in the number of atoms below which
the state becomes delocalized.  In the limit of rapidly varying
NOL`s one can show, using the averaging method, that the system
can be reduced to a nonlinear Schrodinger equation with cubic and
quintic nonlinearities for which the existence of a delocalizing
transition is known. For binary BEC mixtures the same method leads
to a coupled system of cubic-quintic NLS equations for which
delocalizing transitions are also expected to exist. At the
transition point the localized state becomes spatially more
extended and displays properties similar to Townes solitons of the
1D quintic NLS system or of the 2D NLS equation with cubic
nonlinearity. For broad soliton states, i.e. when the soliton
width $l_s$ becomes much larger than the periodicity scale $l_p$,
we can consider the expansion $u_i = U_i + \epsilon u_{1,i} +
\epsilon^2 u_{2,i}+... $, with $\epsilon \sim l_p/l_s << 1$. At
the leading order $1/k^2$ we obtain
\begin{equation}
u_{1,i} = \frac{1}{4}\cos(2x)(\gamma_i |U_i|^2  + g_1 |U_j|^2
)U_i.
\end{equation}
Substituting into Eq.~(\ref{GP}) and averaging over rapid
oscillations we get for the slowly varying functions $U_i$ the
following coupled system with cubic-quintic interactions
{\small
\begin{eqnarray}
iU_{i,t} + U_{i,xx} - \gamma_{i0} |U_{i}|^2 U_i - g_{0}|U_{j}|^2
U_{i} +
\frac{3\gamma_1^2}{8} |U_i|^4 U_i + \nonumber\\
\frac{g_1}{4}(2\gamma_i + g_1)|U_i|^2 |U_j|^2 U_i +
\frac{g_1}{8}(2\gamma_j + g_1)|U_j|^4 U_i = 0.
\end{eqnarray}
}
\begin{figure}
\includegraphics[width=7.6cm,height=7.cm,angle=0,clip]{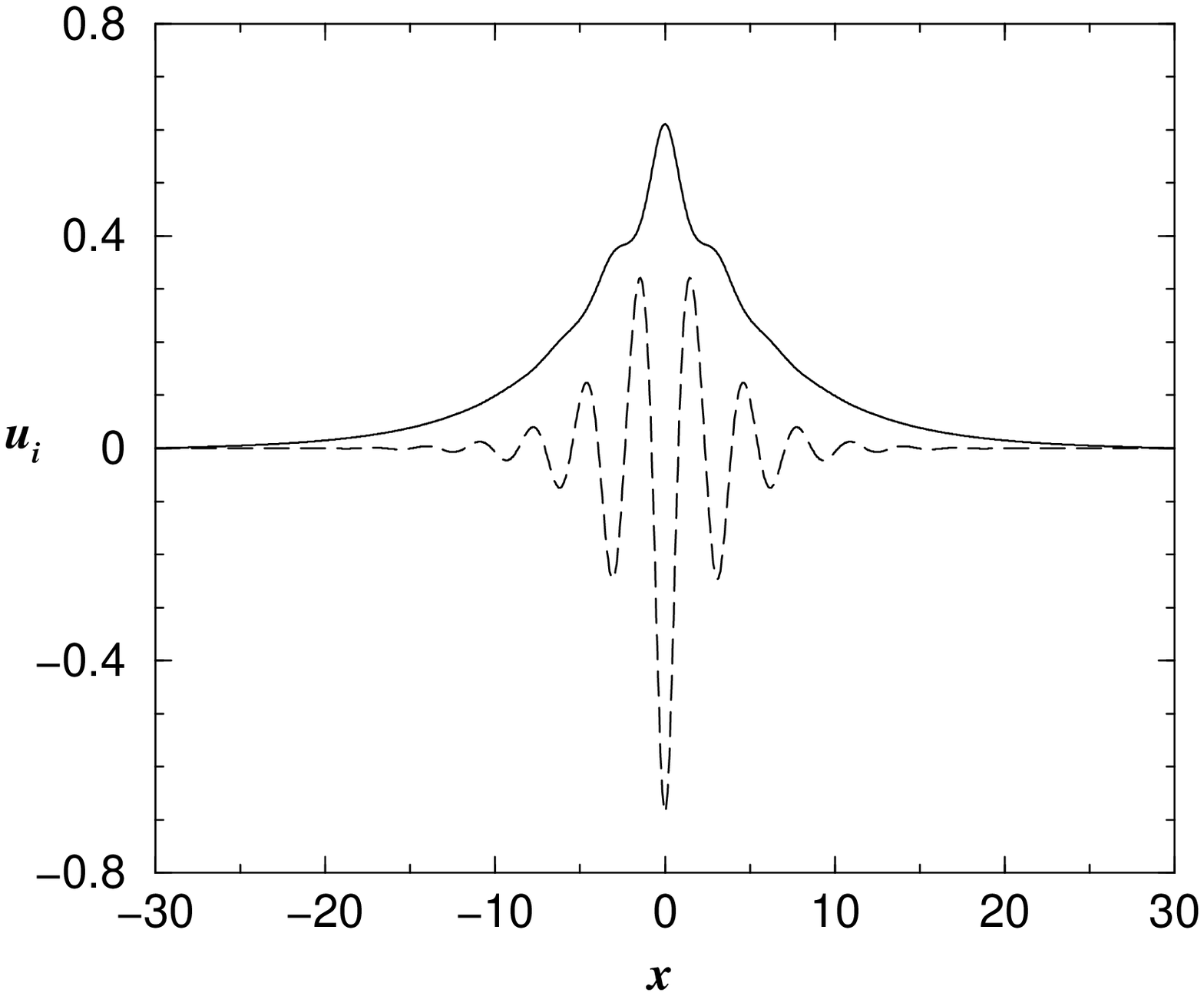}
\includegraphics[width=7.cm,height=7.cm,angle=0,clip]{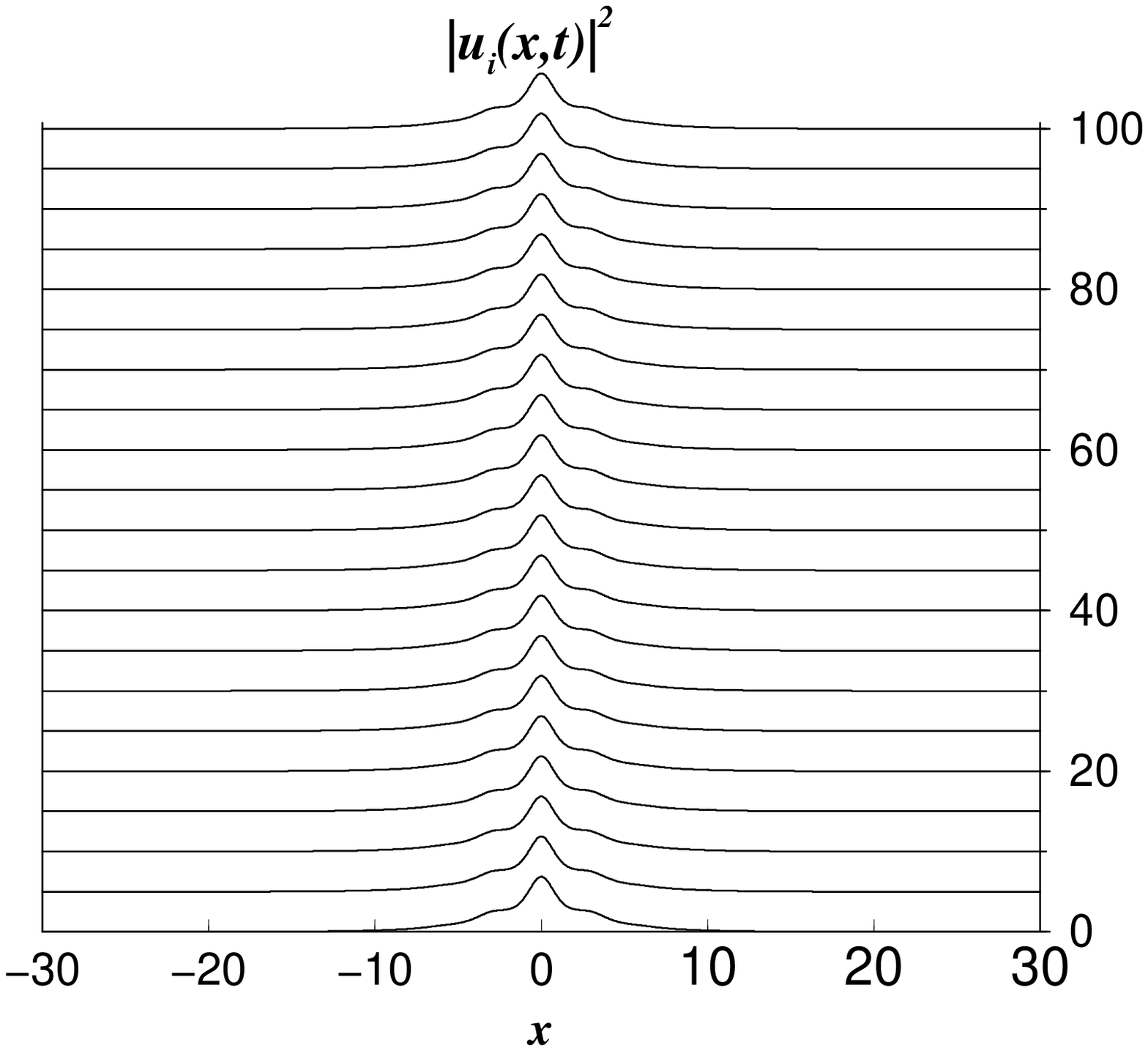}
\vspace{-0.2cm}
 \caption{Top panel. Unstable OS mode of Townes soliton type
 (continuous curve) of Eq.~(\ref{GP}) at the critical strength of the
 inter-species NOL $g_1=-1.3389$ for a delocalizing transition to occur.
 The mode has equal critical number of atoms $N_1=N_2=N_c=2$ with the same
 chemical potential $\mu_1=\mu_2=-0.03498$ and same profiles in both components.
 The dashed line represents the corresponding effective potential $V^{eff}_1=V^{eff}_2$
 in Eqs.~(\ref{self1}) and (\ref{self2}). Other parameters are fixed as
 $\gamma_{10}=-1, \gamma_{20}=-1$, $\gamma_1=\gamma_2=-0.5$, $g_0=1$.
 Bottom panel. Time evolution of the Townes soliton mode in the
 top panel as obtained from Eq.~(\ref{GP}).}
 \label{fig11ms}
\end{figure}
When $\gamma_{i0} = - g_0$, we obtain a system of coupled quintic
NLS equations. For the symmetric case $U_1 = U_2 = U$,  the system
reduces to the quintic NLS equation

\begin{equation}
iU_t + U_{xx} + \chi |U|^4 U=0, \label{qNLS}
\end{equation}
with
\begin{equation}\chi = \frac{3}{8}\gamma_1(\gamma_1 +
g_1) + \frac{g_1}{8}(\gamma_1 + 3g_1 + 2\gamma_2 ).
\end{equation}
The Townes soliton solution of Eq. (\ref{qNLS}) is
\begin{equation}
U(\mu,x) = e^{i\mu
t}(\frac{3\mu}{\chi})^{1/4}\frac{1}{\cosh^{1/2}(2\sqrt{\mu}x)}.
\end{equation}
with  norm given by
\begin{equation}
N_c = \int_{-\infty}^{\infty} |U|^2 dx =
\frac{\pi}{2}\sqrt{\frac{3}{\chi}}.
\end{equation}
This solution behaves as a separatrix between  collapsing and
decaying solutions of the quintic NLSE. Here $dN/d\mu =0$ and the
VK criterion gives marginal stability. The total Hamiltonian is
equal zero on this solution $H(U_{T}) =0$.
For example, for parameters values $g_0 = -1, g_1 =-1.3389,
\gamma_1 = \gamma_2 = -0.5$, we obtain the critical number $N_c =
2.41$. A comparison with the numerical results in
Fig.~\ref{fig11ms} shows that the averaged NLS quintic equation
overestimates the critical number $N_c$ by about $20$
percent(notice that for the same parameters values we have  $N_c =
2$ in Fig.~\ref{fig11ms}). From this we conclude that the quintic
NLS can be used only as a qualitative model for the delocalizing
transitions of two component BEC in NOL. In the following we shall
investigate Townes solitons and delocalizing transitions  by
recurring to numerical methods. Delocalizing transitions in binary
BEC mixtures with NOL and in coupled NLS equations with
cubic-quintic nonlinearities have not been previously
investigated.

To show the existence of this phenomenon in a binary BEC mixture
in a NOL we vary in time the parameter $g_1$ characterizing the
intra-species interaction while keeping fixed the inter-species
NOLs to which the two component are subjected. Starting from a
given value of ${g_1}_0$, for which a stable OS-OS mode exists, we
adiabatically decrease $g_1$ to a value ${g_1}_0-\Delta g_1$ and
then increase it back to the original value.

\begin{figure}
\includegraphics[width=8.5cm,height=6.cm,angle=0,clip]{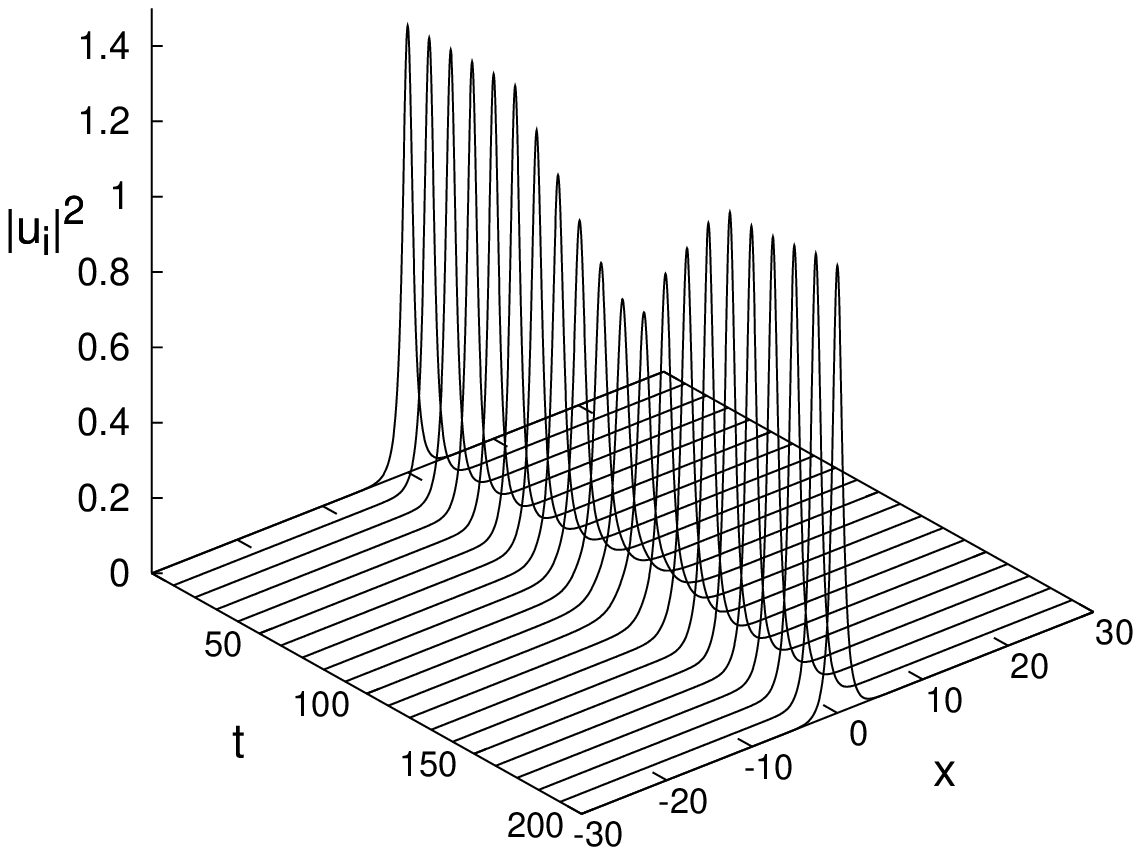}
\includegraphics[width=8.5cm,height=6.cm,angle=0,clip]{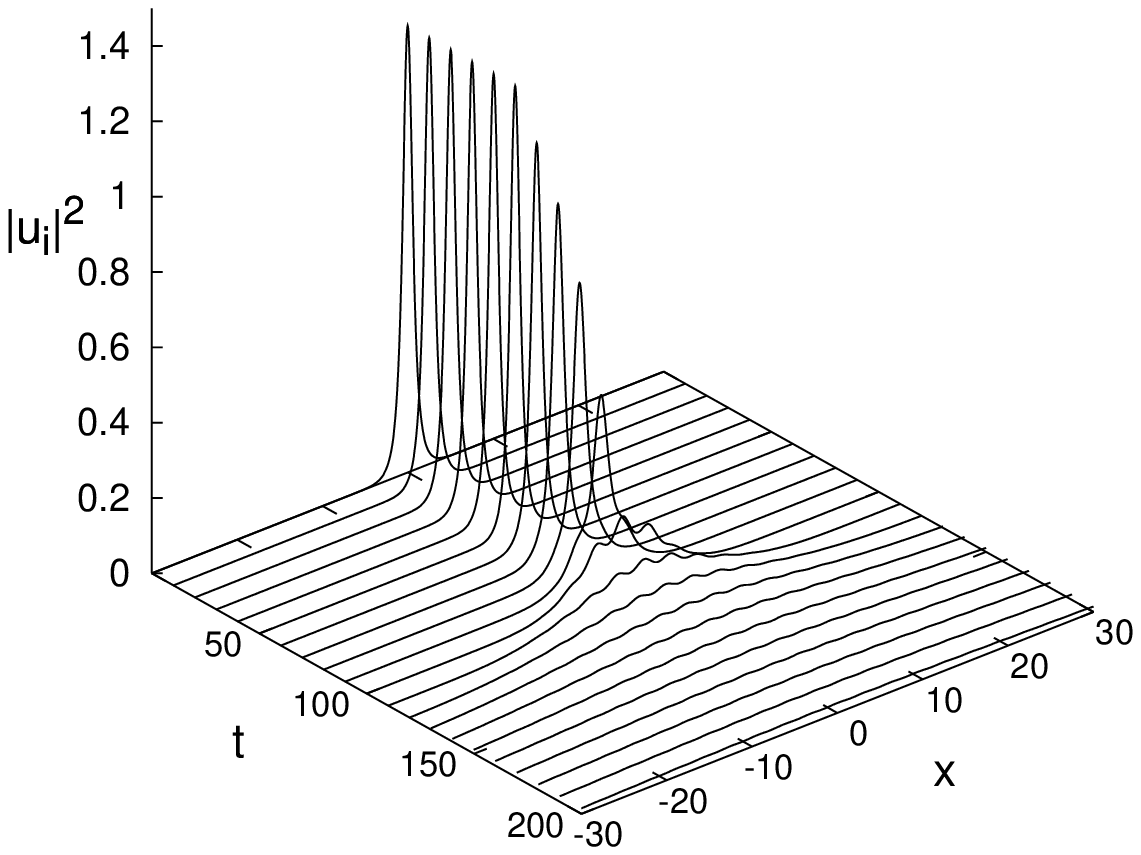}
 \caption{Delocalizing transition of an OS-OS mode of a binary BEC mixture
 with equal number of atoms $N_1=2, N_2=2$. The top panel shows the
 time evolution of the mode when the inter-species parameter $g_1$ is varied
 according to Eq.~(\ref{g1}) with $t_1=50$, $t_2=150$, $\Delta
 g_1=0.25$. The bottom panel shows the
 same evolution but for the case $\Delta g_1=0.34$. Other parameters are fixed as
 $\gamma_{10}=-1, \gamma_{20}=-1$, $\gamma_1=\gamma_2=-0.5$,
$g_0=1$, ${g_1}_0=-1.5$. The chemical potentials of the initial
states at $t=0$ are $\mu_1=\mu_2=-0.5401$.} \label{Fig10}
\end{figure}

In absence of delocalizing transitions the state will restore to
its original form for any decrement $\Delta g_1$, while in
presence of a delocalizing transition a threshold value for
$\Delta g_1$ will appear above which the state becomes
irreversibly delocalized (it cannot be restored to its original
form). In Fig.~\ref{Fig10}
 we show the time evolution of an OS-OS symmetric
state  with an equal number of atoms in the two components, during
a variation of the inter-species interaction in time according of
the form
\begin{equation}
g_1(t)= {g_1}_0 \{1-\Delta g_1 \cos[\pi \frac{t  -\frac12
(t_1+t_2)}{t_2-t_1}]\}. \label{g1}
\end{equation}
From the top panel of this figure it is clear that for a small
decrement $\Delta g_1$ the state is able to restore the initial
waveform, while for a larger decrement the state becomes fully
delocalized. In analogy to what has been done for the NLS equation
with periodic potential and quintic nonlinearity \cite{AS05}, one
can characterize the delocalizing transition in terms  of the
unstable states which separate localized modes from extended ones.
For the parameter used in Fig.~\ref{Fig10}, the critical value
in the strength of the NOL for the occurrence of a delocalizing
transition is found to be $N_c=2$. In Fig.~\ref{fig11ms} we show
the existence of an unstable stationary state found in
correspondence of this value, which has properties similar to the
Townes solitons of the quintic 1D NLS or of cubic multidimensional
NLS.  Note that this stationary state corresponds to
the unstable branch presented in the bottom panel of Fig.~\ref{Fig04} 
[see the exact results for $x_0=0$].

\begin{figure}
\includegraphics[width=7.cm,height=7.cm,angle=0,clip]{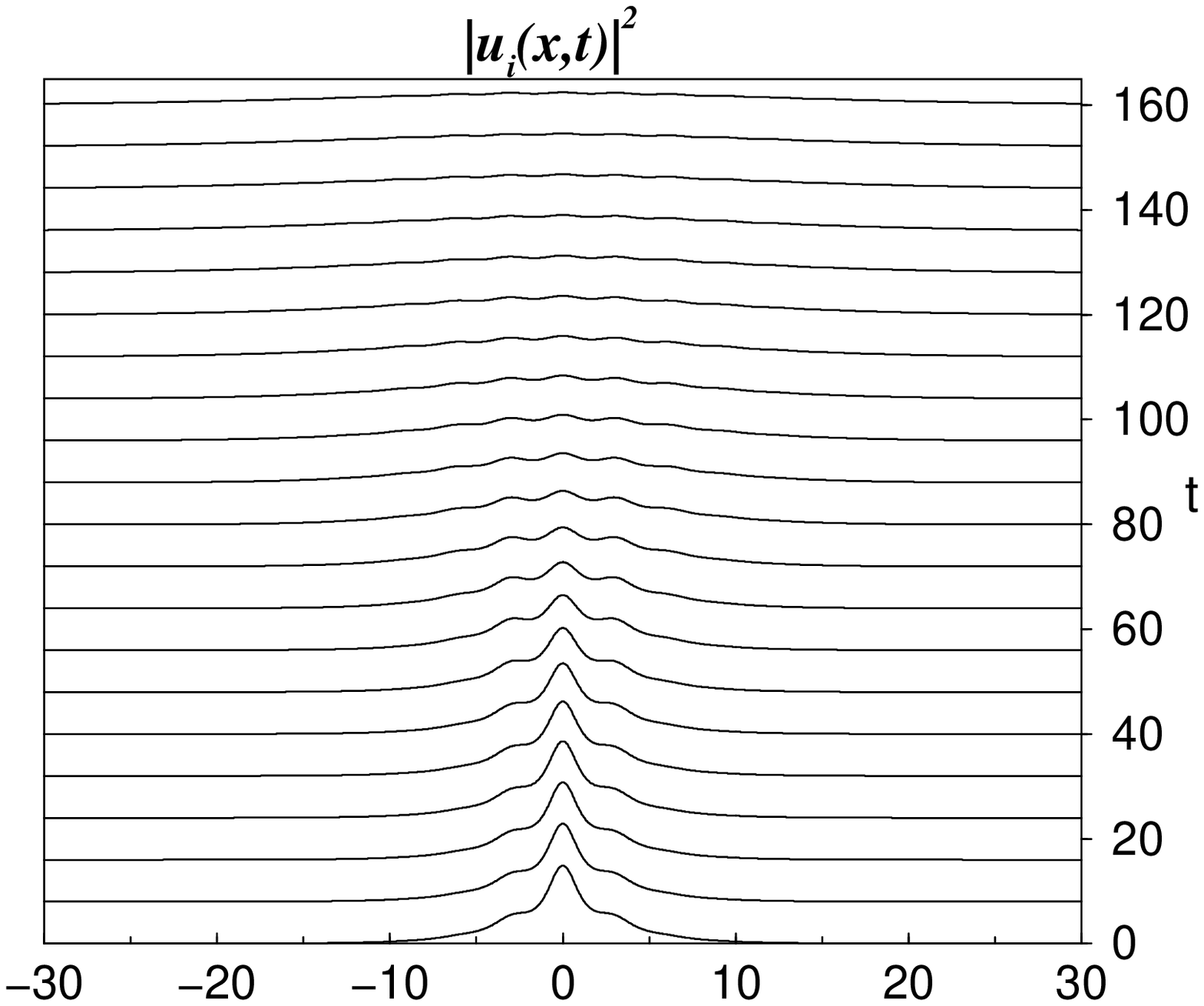}
\includegraphics[width=7.cm,height=7.cm,angle=0,clip]{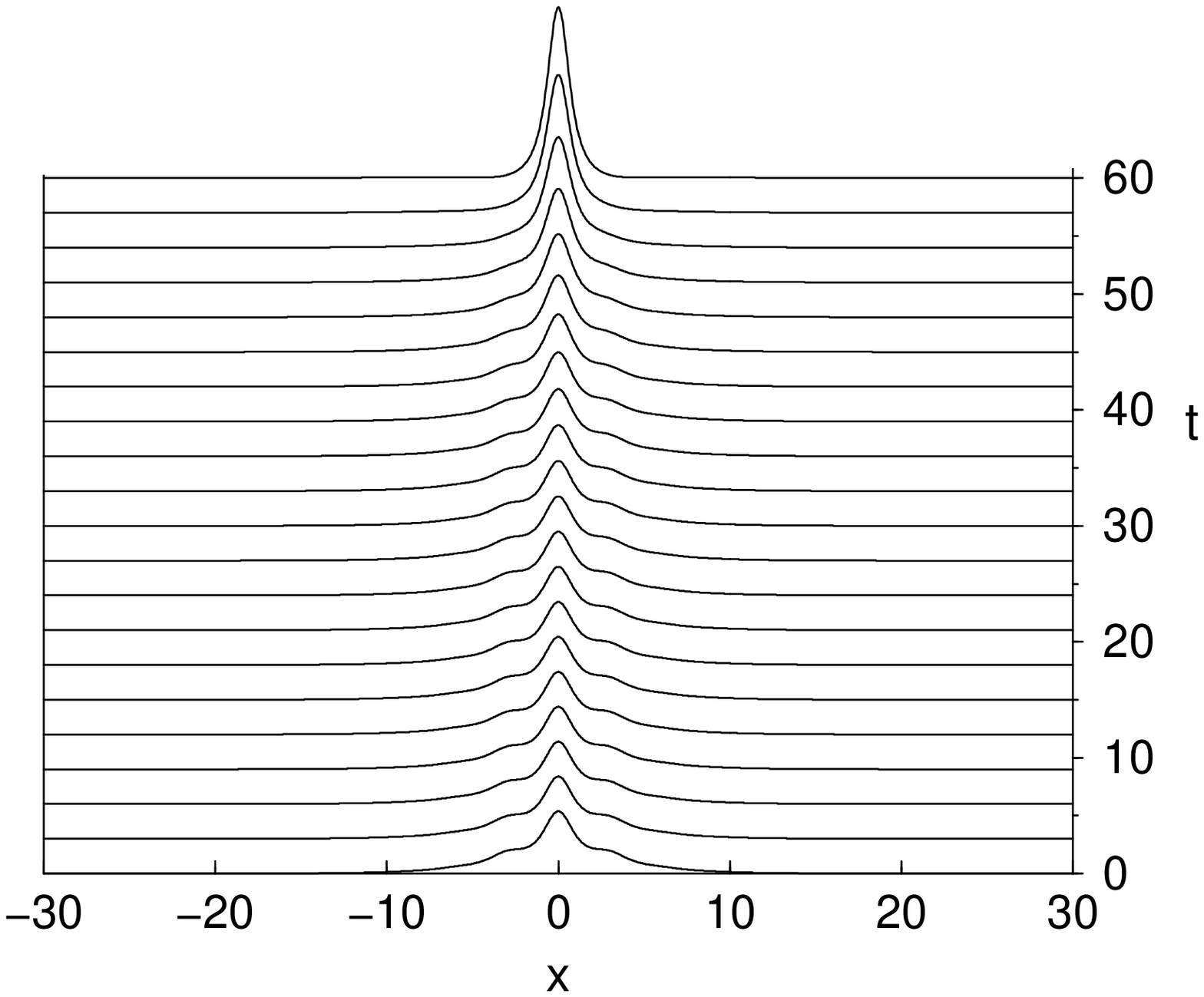}
\caption{Time evolution of the Townes soliton mode in Fig.~\ref{fig11ms} for
 an undercritical $N_{und}=(.999)^2 N_c$ (top panel) and
 overcritical $N_{ov}=(1.001)^2 N_c$ (bottom panel) number of atoms.
  Other parameters are fixed as in Fig.~\ref{fig11ms}.}
 \label{fig12ms}
\end{figure}

From Fig.~\ref{fig12ms}, it is indeed clear  that for slight
undercritical variations of the norm (number of atoms) the state
becomes delocalized, while for slight overcritical variations of
the norm it shrinks to a fully localized mode, resembling the
behavior of Townes solitons. Notice that due to the equal number
of atoms $N_1=N_2$ the modes in the two components have identical
chemical potentials and identical profiles.

\begin{figure}
\centerline{
\includegraphics[width=7.5cm,height=7.5cm,clip]{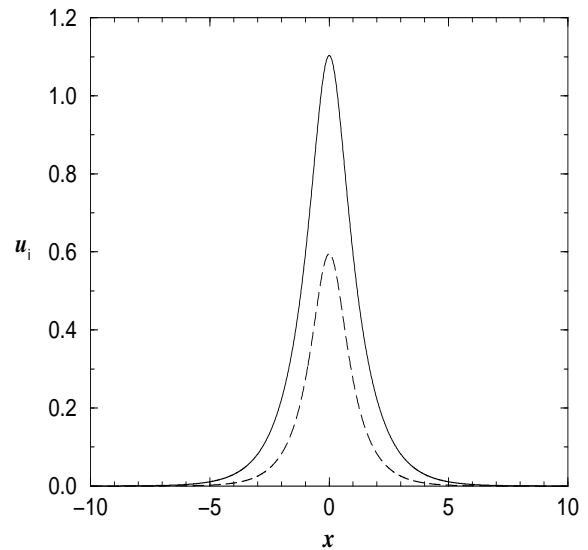}}
 \caption{Onsite symmetric mode of Eq.~(\ref{GP}) with unbalanced number of atoms
 $N_1=2, N_2=0.5$,  and
 for NOL parameters: $\gamma_{10}=-1,
\gamma_{20}=-1$, $\gamma_1=\gamma_2=-0.5$, $g_0=-1$, $g_1=-1.5$.
The continuous (dashed) curve refer to the first (second)
component. The chemical potentials of the modes are $\mu_1=-1.013,
\mu_2=-1.412.$ The dashed line refers to the second component.}
\label{Fig13}
\end{figure}

\begin{figure}
\centerline{
\includegraphics[width=8.5cm,height=8.cm,angle=0,clip]{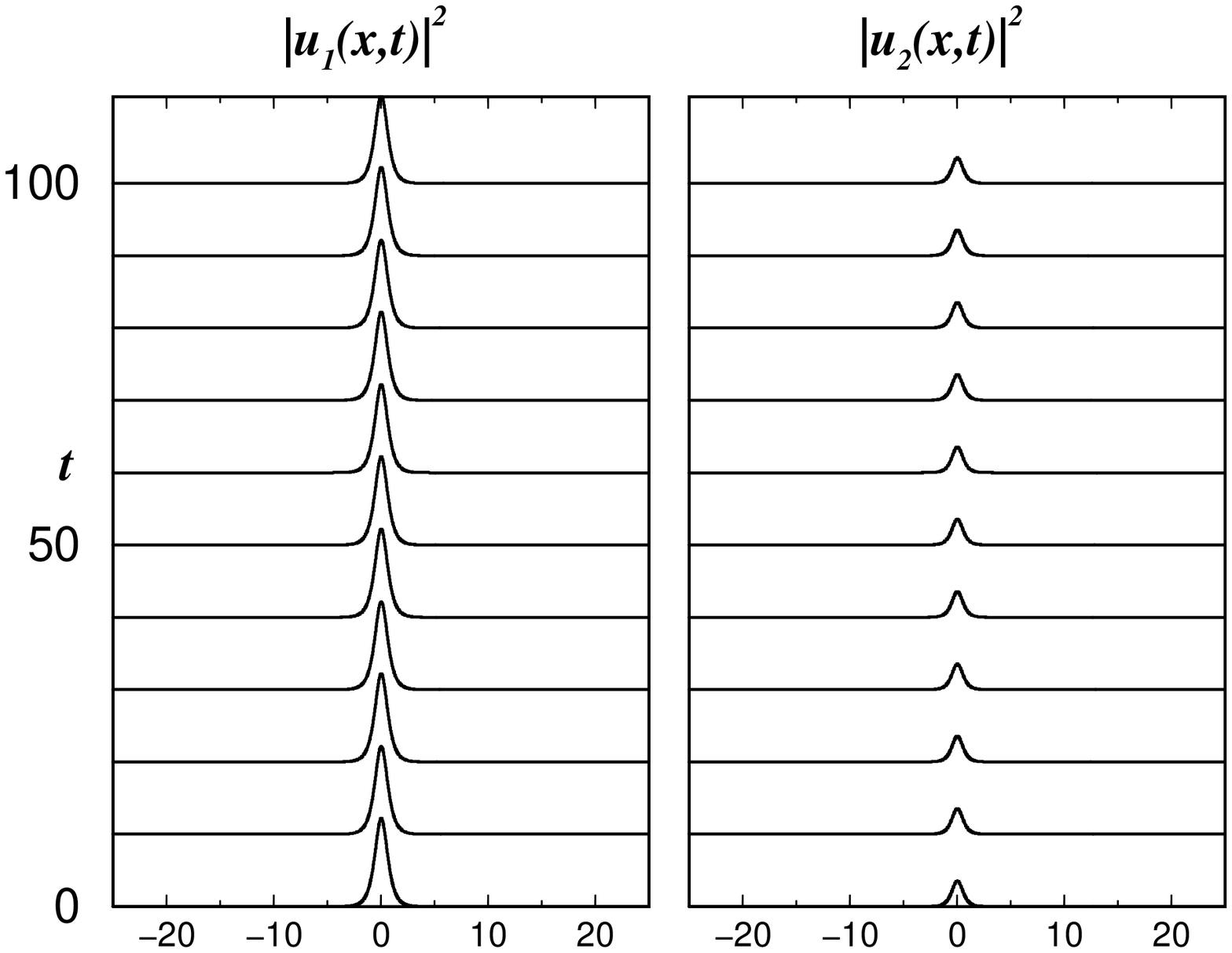}
} \vspace{-1cm}
\centerline{
\includegraphics[width=8.5cm,height=8.cm,angle=0,clip]{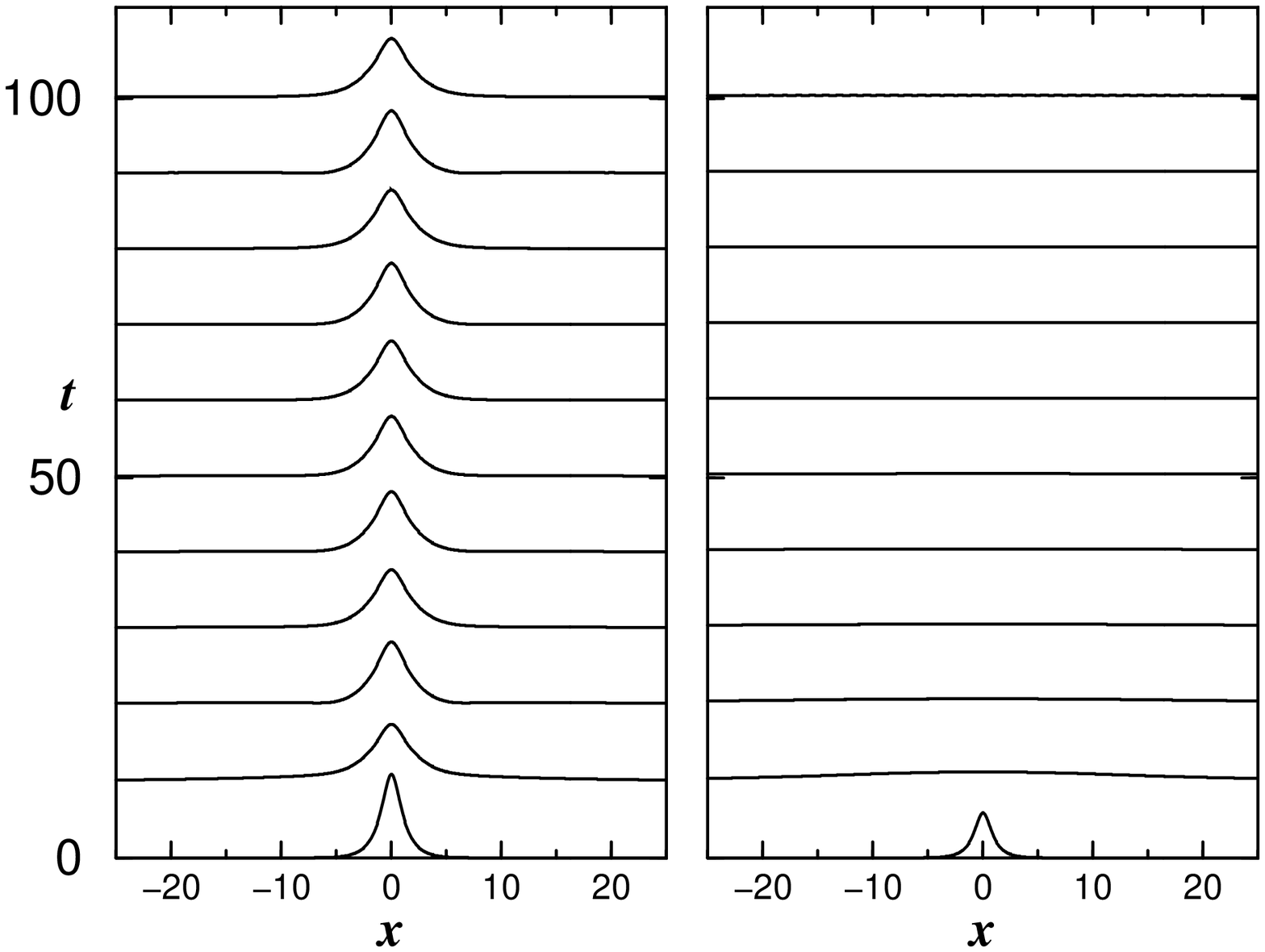}
} \caption{Time evolution of the OS-OS mode in Fig.~\ref{Fig13}
in the presence (top panels) and in the absence (bottom panels) of
the inter-species NOL of strength $g_0=-1, g_1=-1.5$. }
\label{Fig14}
\end{figure}

A delocalizing transition is also observed for OS-OS states with
different number of atoms in the two components. In this case the
system shows a much rich behavior due to the possibility to use
the inter-species interaction to stabilize localized states which
in absence of interaction would be extended over the whole system.
An example of such inter-species induced  localization is given in
Fig.~\ref{Fig14} for an OS-OS symmetric states of Fig.~\ref{Fig13}
with an unbalanced number of atoms  (a large
difference in the number of atoms in the two components).
In particular,  in absence of the inter-species interactions, 
the first component has enough atoms to be above the delocalizing
threshold,  while the second component is taken to be below such a
threshold, so that the state delocalizes in absence of interaction.
From  Fig.~\ref{Fig14} we see, indeed, that the presence of the
inter-species interaction prevents the second component to
delocalize, while in absence of the inter-species interaction the
first component remains localized and the second one delocalizes
in a quite short time. Due to the many parameters of the problem,
a full investigation of the delocalizing transitions of the
fundamental OS-OS mode in binary BEC mixtures with NOL requires
more extensive numerical investigations. We plan to do this in a
separated publication.

\section{Conclusion}
In this paper we have investigated the localized states in
two-component BEC with periodic modulation in space intra-species and
inter-species scattering lengths. The  stability regions are
analyzed using the variational approach and the Vakhitov-Kolokolov
criterion. The symmetry properties (with respect to the NOL) of
the localized modes in each component were considered and their
stability properties investigated. We showed that localized modes
of OS-OS type are always stable and represents the fundamental
ground states of the system in the presence of attractive
interactions. Intersite symmetric modes and  mixed symmetry modes
also exist but they appear to be metastable under GPE time
evolution, decaying into modes of OS-OS-type. The existence
regions in the parameter space  of strongly localized modes
(localized on few cells of the NOL) of fundamental type  were
predicted by mean of the variational ansatz and their stability
properties predicted by the Vakhitov-Kolokolov criterion.
Localized modes on tops of periodic backgrounds and of bright-dark
solitons were also shown to exist in the case of binary mixtures
with opposite interactions in the two components.

In spite of the quasi 1D nature of the problem we showed that
fundamental solitons undergo a delocalizing transition when the
strength of the intersite non linear optical lattice is varied.
This transition was associated with the existence of an unstable
localized solution which  extends on many lattice cells of the the
NOL and  which exhibit a shrinking (decaying) behavior for
slightly overcritical (undercritical) variations in the number of
atoms.

This behavior was shown to exists for fundamental modes both with
equal and unequal   numbers of atoms in the two components.

The existence of the delocalizing transition for the fundamental
modes was inferred  also from a reduced vector GPE obtained by
averaging the original GPE system with respect to the rapid
spatial oscillations introduced by the NOL. The process of
averaging the NOL introduces high order nonlinearities
(cubic-quintic) which make the problem to be effectively
equivalent to an higher dimensional vector GPE system for which
delocalizing transition, in analogy to single component
multidimensional cases, are usually expected.

The study of the delocalizing transition for fundamental
multi-component solitons  in terms of an averaged vector GPE with
higher order nonlinearities, as well as the extension of the above
analysis to the multidimensional case, appear to be  interesting
problems which deserve further investigations.

\section*{Acknowledgments}
FKA and MS wish to thank the Instituto de F\'\i sica Te\'orica,
Universidade Estadual Paulista (UNESP) for hospitality.
For the financial support, which makes possible to realize this collaboration,
we thank Funda\c c\~ao de Amparo \`a Pesquisa do Estado de S\~ao Paulo
(FAPESP).
MS acknowledges partial financial support from the MUR through the
inter-university project PRIN-2005: ``Transport properties of
classical and quantum systems". AG and LT also thank Conselho Nacional de
Desenvolvimento Cient\'\i fico e Tecnol\'ogico (CNPq) for partial
financial support.

\section*{Appendix  - Numerical Approach}

The numerical methods employed in this paper are described in the
following subsections.

\subsection{Self-consistent diagonalization algorithm}

We solve the nonlinear eigenvalue problem in (\ref{GP}) by
treating the nonlinear part in self consistent manner. This
amounts to consider the following linear eigenvalue problem
\begin{eqnarray}
\label{self1} \mu_{1} u_{1} = -\frac{\partial^2 u_{1}}{\partial
x^2} &+& V^{eff}_1 u_{1}, \\
\label{self2} \mu_{2} u_{2} = -\frac{\partial^2 u_{2}}{\partial
x^2} &+& V^{eff}_2 u_{2},
\end{eqnarray}
with the effective potentials defined as $V^{eff}_i=\gamma_i(x)
|u_i|^2 + g(x)|u_{3-i}|^2$, $\;i=1,2$. To solve these eigenvalue
problem we adopt a discrete variable representation \cite{harris}
and  diagonalize the operators $\hat H_i= \hat K+ \hat V^{eff}_i$
in the discrete coordinate space representation $\{x_n= n a \}$,
$n=1,..., N_p$, $a=L/N_p$. Here $\hat K$ denotes the kinetic
energy operator $\hat K=-\frac{\partial^2} {\partial x^2}$,  $L$
is the length of the system and $N_p$ the number of grid points.
By taking as a basis the set of vectors $|x_n>
=(0,...0,1,0,...0)$, $n=1,...,N_p$ and noting that $V^{eff}_i$ is
already diagonal in this basis while $\hat K$ is diagonal in the
momentum representation $\langle k_n | \hat K |k_m \rangle= k_n^2
\delta_{n,m}$, we have that the matrix elements of $\hat H_i$ can
written as
\begin{equation}
\langle x_n| \hat H_i |x_m \rangle= \langle x_n | \hat F^{-1} \hat
K \hat F |x_m \rangle + V^{eff}_i(n a) \delta_{n,m},
\label{matrices}
\end{equation}
where $\hat F |x_n\rangle$  denotes the Fourier (unitary)
transform of the vector $|x_n\rangle$. Standard diagonalization
routines are then used to find eigenvalues (chemical potentials)
and eigenfunctions. The nonlinear eigenvalue is then solved in
self-consistent manner starting from trial wavefunctions $u_{1}$,
$u_{2}$, calculating the effective potentials , solving the
eigenvalue problems (\ref{self1}) by diagonalizing the
correspondig matrices (\ref{matrices}), selecting given
eigenstates as new trial functions, and iterating the procedure
until convergence is
 reached (see Refs.\cite{LP05,Cruz07} for applications to single
 and multicomponent BEC  cases).

\subsection{Relaxation technique}
The method of relaxation technique was used to check the results
obtained with the previous method, to improve their accuracy, and
also to make a complete study on the stability of the solutions.

Stable states are obtained using standard relaxation algorithm in
imaginary time propagation, fixing the normalizations given by
number of atoms of the two species, $N_1$ and $N_2$, and obtaining
the chemical potentials $\mu_1$ and $\mu_2$. For the hyperbolic
(unstable) states we extended to a coupled equation system the method
developed in Ref.\cite{marijana}, scheme C, in which the
idea of ``back renormalization" was used. In this method, it is given
the chemical potential to obtain the number of atoms.

For a coupled system, the scheme C of Ref.~\cite{marijana}
can be generalized, evolving the following equations in
imaginary time:`
{\small
\begin{eqnarray}
-\frac{\partial \varphi}{\partial \tau}&=&
\left(-\frac{\partial^2 }{\partial x^2}
+N_1\beta_1 |\varphi|^2 + N_2\sigma_{12}|\phi|^2-\mu_1\right) \varphi
\label{ap1} \\
-\frac{\partial \phi}{\partial \tau}&=&\left(-\frac{\partial^2 }{\partial x^2}
+N_2\beta_2|\phi|^2 +N_1\sigma_{12}|\varphi|^2-\mu_2
\right)\phi 
\label{ap12}
,\end{eqnarray}}
where we have normalized $\varphi$ and $\phi$ to one, such that $\varphi
\equiv u_1/\sqrt{N_1}$ and $\phi\equiv u_2/\sqrt{N_2}$. $\beta_i$ and $\sigma_{12}$
are given by Eq.~(\ref{gammas}).

In discretized version, the coupled equations (\ref{ap1}) and (\ref{ap12}) takes the form

\begin{eqnarray}
\varphi^{n+1/3}&\leftarrow& \varphi^n+\frac{\Delta\tau}2
\left(\mu_1-\beta_1 N_1^n|\varphi^n|^2 -\sigma_{12}N_2^n|\phi^n|^2
\right)\varphi^n\nonumber\\
\varphi^{n+2/3}&\leftarrow& O_{CN}\varphi^{n+1/3}
\nonumber\\
\varphi^{n+1}&\leftarrow& \varphi^{n+2/3}+\frac{\Delta\tau}2
\left(\mu_1-\beta_1 N_1^n|\varphi^n|^2-\sigma_{12} N_2^n|\phi^n|^2
\right)\varphi^n \nonumber
\end{eqnarray}
\begin{eqnarray}
\phi^{n+1/3}&\leftarrow& \phi^n+\frac{\Delta\tau}2
\left(\mu_2-\beta_2N_2^n|\phi^n|^2-\sigma_{12}N_1^n|\varphi^n|^2
\right)\phi^n\nonumber\\
\varphi^{n+2/3}&\leftarrow& O_{CN}\varphi^{n+1/3}
\nonumber\\
\phi^{n+2/3}&\leftarrow& \phi^{n+2/3}+\frac{\Delta\tau}2
\left(\mu_2-\beta_2N_2^n|\phi^n|^2-\sigma_{12}N_1^n|\varphi^n|^2
\right)\phi^n \nonumber
\end{eqnarray}
\begin{eqnarray}
N_1^{n+1}&\leftarrow&
\frac{N_1^n}{\displaystyle\int dx |\phi^{n+1}|^2},
\nonumber \\
N_2^{n+1}&\leftarrow&\frac{N_2^n}
{\displaystyle\int dx |\varphi^{n+1}|^2},
\nonumber \\
\varphi^{n+1}&\leftarrow&\frac{\varphi^{n+1}}
{\displaystyle\int dx |\varphi^{n+1}|^2},
\nonumber\\
\phi^{n+1}&\leftarrow&\frac{\phi^{n+1}}
{\displaystyle\int dx |\phi^{n+1}|^2},
\nonumber
\end{eqnarray}
where the superscripts ($n$, $n+1$, etc) refer to time steps.
$O_{CN}$ is the Crank-Nicolson evolution operation corresponding to
$-{\partial^2 }/{\partial x^2}$. Note that, in this
coupled system the back renormalization (of $N_1^{n+1}$ and $N_2^{n+1}$)
is done by exchanging the corresponding wavefunctions
(as $N_1$ is associated to $\varphi$ and $N_2$ to $\phi$). This procedure
is required for stability, as verified in numerical tests.

The excited states IS-OS and IS-IS depicted in Fig.~\ref{Fig06} can be obtained by
relaxing Eqs.~(\ref{ap1})-(\ref{ap12}) for $x\ge 0$ and imposing the Von Neumann boundary conditions
in the origin, i.e., at $x=0$, $\partial
\varphi/dx=0$ and $\partial \phi/\partial x=0$.
The present relaxation algorithms are unable to find the state shown in
Fig.~\ref{Fig08}, which was obtained by the approach given in subsection A.

As compared to the scheme shown in subsection A, the advantage of relaxation methods 
relies on the possibility of generalization to higher dimensions with few computational resources.

\end{document}